\documentclass[11pt,aps,prd,nofootinbib,superscriptaddress,preprintnumbers]{revtex4}
\usepackage{graphicx,epsfig}
\usepackage{float}
\usepackage{color}

\newcommand{\lsim}{\raisebox{-0.13cm}{~\shortstack{$<$ \\[-0.07cm] $\sim$}}~}
\newcommand{\gsim}{\raisebox{-0.13cm}{~\shortstack{$>$ \\[-0.07cm] $\sim$}}~}

\begin{document}
{\small
\begin{flushright}
CNU-HEP-15-05 \\
FTUV-2015-07-29\\
IFIC/15-48 \\
\today
\end{flushright} }
\title{Flavour-changing Higgs decays into bottom and 
strange quarks \\[-1mm] in supersymmetry}
\author{G.~Barenboim}\email{Gabriela.Barenboim@uv.es}
\author{C.~Bosch}\email{Cristian.Bosch@uv.es}
\affiliation{Departament de F\'{\i}sica Te\`orica and IFIC, Universitat de 
Val\`encia-CSIC, E-46100, Burjassot, Spain.}
\author{J.S.~Lee}\email{jslee@jnu.ac.kr}
\affiliation{Department of Physics, Chonnam National University, 300 Yongbong-dong, Buk-gu, Gwangju, 500-757, Republic of Korea}
\author{M.L.~L\'opez-Ib\'a\~nez}\email{M.Luisa.Lopez-Ibanez@uv.es}
\author{O. Vives}\email{Oscar.Vives@uv.es}
\affiliation{Departament de F\'{\i}sica Te\`orica and IFIC, Universitat de 
Val\`encia-CSIC, E-46100, Burjassot, Spain.}
\bigskip
\begin{abstract}
In this work, we explore the flavour changing decays $H_i \to b s$ in a general supersymmetric scenario. In these models, the
flavour changing
decays arise at loop-level but, originating from a dimension-four operator, do not decouple and may provide a first sign of new physics for heavy masses beyond collider reach. In the framework of the
minimal supersymmetric extension of the Standard Model (MSSM),
we find that
the largest branching ratio of the lightest Higgs ($H_1$)
is ${\cal O}(10^{-6})$
after imposing present experimental constraints, 
while heavy Higgs states may still present branching ratios ${\cal O}(10^{-3})$. 
In a more general supersymmetric scenario, where additional Higgs states may modify the Higgs mixings, the branching ratio BR($H_1 \to b s$) can reach values 
${\cal O}(10^{-4})$ , while heavy Higgses still remain at
${\cal O}(10^{-3})$. Although these values are clearly out of reach for the LHC, a full study in a linear collider environment could be worth.
\end{abstract}
\maketitle

\newpage
\section{Introduction}

Since the discovery of a scalar boson with mass $\sim$126 GeV, at the LHC in 2012 \cite{Aad:2012tfa,Chatrchyan:2012ufa},
the Standard Model (SM) picture may have been completed. Indeed, if this scalar particle corresponds to the SM Higgs boson, the SM could be the correct description of nature up to scales close to the Planck mass. So far, all the experimental evidence seems to be pointing in the direction of confirming that it is really the missing piece of the SM puzzle. Nevertheless the exploration of 
 features of this particle
is just beginning and further studies are needed to confirm its identity. 

From now on, our efforts to probe the SM and to search for physics beyond it may follow two complementary paths: i) push the energy frontier in the search for new particles and interactions and ii) increase the precision on the couplings of the first (so far) fundamental scalar ever discovered. Indirect searches, which would include the latter path, searches of rare processes or higher order corrections to low energy couplings, have been very successful in the past and have led to the discovery of new particles
such as the third generation quarks. They have also been instrumental in exploring the scale of new physics beyond the collider reach.

In the case of the Higgs boson, the study of its couplings can be the way to go.
Scrutinizing
non-standard Higgs couplings is a way to test the presence of additional scalar
bosons even when their direct production is closed. 
In the models beyond the SM in which there exists
more than one Higgs doublet, as in a two Higgs doublet model 
(2HDM)\cite{Branco:2011iw} or the MSSM \cite{Nilles:1983ge,Haber:1984rc}, 
the couplings of the candidate for the discovered scalar boson
may not be flavour diagonal and thus it can have flavour changing decays \cite{Hamzaoui:1998nu,Choudhury:1998ze,Babu:1999hn,Chankowski:2000ng,Isidori:2001fv,Bobeth:2002ch,Buras:2002wq,Buras:2002vd,Paradisi:2005tk,Arhrib:2006vy,Buras:2010mh,Crivellin:2010er,Crivellin:2011jt,Crivellin:2012zz,Goudelis:2011un,Gabrielli:2011yn,Blankenburg:2012ex,Arana-Catania:2013xma}. Furthermore, these flavour changing
couplings are dimensionless and therefore the effects of
additional heavy particles may
not decouple even
when the masses of such particles are taken to infinity, 
providing a unique opportunity to find 
an assuring indirect evidence for such a high scale
non-easily accessible in the near future.

Even more, as we have no clue of the new energy scale (if any) associated with the  new particles and interactions we are looking for, considering rare Higgs
decays is a wise way to go. Thus, our proposal consists of searching for the Flavour-Changing (FC)  Higgs decays. In the SM these FC decays do not exist and therefore their presence would undoubtedly signal the presence of new physics beyond the SM.

As mentioned before, in the past, rare decays, including processes like $b\rightarrow s\,\gamma$ and $B_s\rightarrow\mu^{+}\,\mu^{-}$, have been extensively used to search for new physics. Their precise experimental measurements were useful for the exploration of the parameter space of different SM extensions. Likewise, 
FC Higgs decays 
are very useful to search for new physics since they are
present in almost any extension of the SM containing additional scalars 
that mix with the lighter Higgs. 
For instance they are unavoidable in type-II 2HDMs, like the MSSM, pseudo-dilaton models \cite{Goldberger:2007zk,Fan:2008jk}, models with extra dimensions \cite{Casagrande:2008hr,Azatov:2009na} or composite Higgs models \cite{Agashe:2009di}. To be specific,  in this work we will explore a generic supersymmetric scenario, as well as 
the MSSM framework, both in the presence of 
non-minimal flavour structures.

Our analysis will be focused on the process $H\rightarrow b\,s$, since one would
expect on general grounds that, among FC Higgs decays, 
those involving third generation particles whose Yukawa couplings are larger are
the most experimentally accessible. Besides, loop-induced FC processes in the 
quark sector are typically larger by a factor $\alpha_3^2/\alpha_2^2$ (where $\alpha_3$ and $\alpha_2$ are the coupling constants associated to the groups $SU(3)$ and $SU(2)$ respectively) for similar 
flavour changing entries in the lepton sector.

The main goal in this analysis is
to find out the largest FC branching ratios for the different Higgs states attainable in a general supersymmetric scenario. 
As we will show, in the MSSM, 
the decoupling of the heavy Higgses, enforced by present constraints, makes the FC branching ratio (BR) of the observed light-Higgs to be below the level of $10^{-6}$, 
while heavy Higgs states could reach $\sim 10^ {-3}$. 
In a more general supersymmetric standard model,
the light Higgs BR could still reach values of $\sim 10^{-4}$.
Thus, these rare decays are very challenging for the LHC but they can be searched for at lepton colliders as the International Linear Collider (ILC). As will be shown,  the branching ratio can
reach a level of $10^{-3}$ under special circumstances in the generic SUSY case.
These branching ratios could be reached at a linear collider. In fact, we can see that already at LEP, a limit of BR$(Z\to b \bar s) < O(10^{-3})$ was obtained with only $O(10^6)$ $Z$-bosons in Ref.~\cite{Fuster:1999dj}. We can expect the larger statistics and improved experimental techniques to improve these limits (see also \cite{Eilam:2002as}). In the case of the Higgs decays, we can produce between $O(10^5)$ and  $O(10^6)$ Higgs bosons for $m_H=125$ GeV \cite{Asner:2013psa}, and therefore we can expect similar values for the lightest Higgs branching ratio. 
Thus, FC Higgs decays may provide an indirect hint for the existence of 
new physics at higher energies even when these higher scales are beyond  
the LHC or ILC reach. 

This paper is organized as follows: in section \ref{sec:fc}, we introduce the framework in which the analysis will be carried out. Some compelling variations of it will also be addressed because of their interest when trying to observe FC processes. 
Still in this section, the theoretical expressions for 
the FC Higgs decays into down-type quarks will be provided. 
Section \ref{sec:exconstr} summarizes the latest experimental data corresponding
to collider probes and indirect bounds from low-energy experiments. Finally,
section \ref{sec:numanalysis} collects the main features of the numerical
analysis, concluding in section \ref{sec:conclusions} with the main results.

\section{Higgs Flavour Changing in the MSSM} \label{sec:fc}

Our
analysis is performed within a generic CP-violating MSSM framework
and its extensions, in which
the minimal Higgs sector is 
a type-II 2HDM, {\it i.e.} one of the two scalar doublets couples only to the 
up-type quarks at tree-level while the other couples only to down-type quarks. When electroweak symmetry breaking (EWSB) occurs, 
the neutral components of the two Higgs fields acquire vacuum expectation values
(VEVs) and five physical Higgs states appear: three with neutral electric 
charge and two charged bosons. The two Higgs doublets can be parameterized as
  \begin{equation}
  \Phi_1\:=\:\left(\begin{array}{c}
  \Phi_1^0 \\
  \phi_1^-
  \end{array}\right)\,,
  \hspace{2.cm}
  \Phi_2\:=\: e^{i\xi}\,\left(\begin{array}{c}
  \phi_2^+ \\
  \Phi_2^0
  \end{array}\right)
  \label{eq:eqn201}
  \end{equation}
where $\Phi_i^0\:=\:\frac{1}{\sqrt{2}}\left(v_i+\phi_i+i\,a_i\right)$,
with $v_1=v\cos\beta$, $v_2=v\sin\beta$
and $v\simeq 246$ GeV. 
At tree level the mass eigenstates ($H_i$) are CP eigenstates, but this situation changes once loop corrections are taken into account \cite{Pilaftsis:1998pe,Pilaftsis:1998dd,Demir:1999zb,Pilaftsis:1999qt,Carena:2000yi,Choi:2000wz,Carena:2001fw}. 
In the MSSM, 
the possibility of having CP-violating phases increases due to the growing 
number of complex parameters in the 
{so-called soft SUSY breaking terms},
and indeed these CP phases contribute at loop-level to Higgs masses and mixings.
Consequently, weak-state fields ($\phi_{1,2}$ and $a$) give rise to CP-mixed mass
eigenstates ($H_i$) and these states are related through a unitary
transformation represented by the $3\times3$ orthogonal
mixing matrix $\mathcal{O}$:
  \begin{eqnarray}
  \phi_{1}\:=\:\mathcal{O}_{1i}\,H_i\,, \hspace{1.cm} &
\phi_2\:=\:\mathcal{O}_{2i}\,H_i\,, \hspace{1.cm} &
a\:=\:\mathcal{O}_{3i}\,H_i\,.
  \label{eq:eqn204}
  \end{eqnarray}
The mass eigenvalues will be obtained by means of diagonalising the mass 
squared matrix:
\vspace{0.1cm}
  \begin{equation}
  \mathcal{O}^T\,\cdot\,\mathcal{M}^2_{H}\,\cdot\,\mathcal{O}\:=\:\mathrm{Diag}\left(m_{H_1}^2,m_{H_2}^2,m_{H_3}^2\right)
  \label{eq:eqn205}
  \end{equation}
To study Higgs flavour changing decays, it is helpful to introduce a convenient parametrization of the Higgs mixings. During the analysis, we will use
  \begin{equation}
  \delta_1\,\equiv\,\left(\frac{\mathcal{O}_{11}}{\cos\beta}-\frac{\mathcal{O}_{21}}{\sin\beta}\right)
\hspace{2.cm} \eta_1\,\equiv\,\frac{\mathcal{O}_{31}}{\sin\beta\cos\beta}\,, 
  \label{eq:eqn206}
  \end{equation} 
where $\delta_1$ quantifies the distance of the lightest Higgs mixings from $\cos \beta$ and $\sin \beta$ and $\eta_1$ is directly related to the pseudoscalar content of $H_1$. 

In our analysis below, we will distinguish two different situations in regard to the Higgs mixing: 
\begin{itemize} 
\item Full MSSM framework: Here, we consider the usual MSSM Higgs potential
\cite{Haber:1984rc} which breaks the electroweak symmetry radiatively. 
The minimization of this potential gives us the Higgs masses and mixings. 

Using $\mathcal{O}_{11}^2+\mathcal{O}_{21}^2+\mathcal{O}_{31}^2=1$, one may
express the mixing angles $\mathcal{O}_{\alpha 1}$ in terms of
$\delta_1$ and $\eta_1$ as follows
\begin{eqnarray}
\mathcal{O}_{11}&=&\cos\beta\left[
\sqrt{1-(\delta_1^2+\eta_1^2)\cos^2\beta\sin^2\beta}
+\delta_1\sin^2\beta \right]\,, \nonumber \\
\mathcal{O}_{21}&=&\sin\beta\left[
\sqrt{1-(\delta_1^2+\eta_1^2)\cos^2\beta\sin^2\beta}
-\delta_1\cos^2\beta \right]\,, \nonumber \\
\mathcal{O}_{31}&=&\eta_1\,\cos\beta\sin\beta\,.
\end{eqnarray}
Then, the coupling of the lightest Higgs to a pair of massive 
vector bosons is given by
\begin{equation}
g_{_{H_1VV}}=\cos\beta\,\mathcal{O}_{11}+\sin\beta\,\mathcal{O}_{21}
=\sqrt{1-(\delta_1^2+\eta_1^2)\cos^2\beta\sin^2\beta}\,.
\label{eq:eqn207}
\end{equation}
The current LHC Higgs data constrain $g_{_{H_1VV}}$ to be close to its SM value, 
$g_{{H_1VV}}=1$. 
In fact, the present best-fit values and uncertainties are 
$\kappa_V =g_{_{H_1VV}} = 1.15 \pm 0.08$ 
if we assume there is no change in the Higgs total width, {\it i.e.} $\kappa_H^2 =\Gamma_H/\Gamma_H^{\rm SM} =1$, and
$\kappa_{VV} = \kappa_V\cdot \kappa_V/\kappa_H=1.28^{+0.16}_{-0.15}$ 
if we allow for a change in the total 
decay width~\cite{ATLAS-CONF-2014-009}.
At present, the errors are still large, but requiring, for example, $g_{{H_1VV}} \gsim 0.9$,
one needs to have
$(\delta_1^2+\eta_1^2)\cos^2\beta\sin^2\beta \lsim 0.2$. 
As we will show later,  
$\mathrm{BR}\left( H_1 \rightarrow \bar{b}s+\bar{s}b\right)$
is directly proportional to the quantity $(\delta_1^2+\eta_1^2)$
which can be larger for larger $\tan\beta$ values
while satisfying this constraint.
On the other hand, large $\tan\beta$ values are constrained by the
$\Delta B=1$ and $\Delta B=2$ processes such as
$b \to s \gamma$, $B_s^0\to \mu^+ \mu^-$, $B_s^0-\bar{B}_s^0$ mixing, etc.
Taking into account all these constraint, we find that,
$\mathrm{BR}\left( H_1 \rightarrow \bar{b}s+\bar{s}b\right)$
can be as large as $10^{-6}$ in an MSSM framework. 

\item Generic supersymmetric SM: Given that no signs of supersymmetry have been found so far in collider experiments, and taking into account the strong constraints on the parameter space of minimal models, it is interesting  to consider more general models.    
In fact, the situation could be different if we consider SUSY models
beyond the MSSM which contain additional Higgs states. In this case, 
the Higgs mass eigenstates $H_i$ are given by
\begin{equation}
H_i = \sum_{\alpha=1,2}\mathcal{O}_{\alpha i}\,\phi_\alpha
    + \mathcal{O}_{3 i}\,a
    + \sum_{\beta\geq 4}\mathcal{O}_{\beta i}\,\varphi_\beta
\end{equation}
where $\varphi_\beta$ represent the additional 
CP-even and CP-odd Higgs states which can be charged or neutral under SU(2)$_L$.
We note that only the SU(2)$_L$-charged CP-even states contribute to the
tree-level $g_{_{H_iVV}}$ couplings and,  due to the additional states, we generically have
$\mathcal{O}_{11}^2+\mathcal{O}_{21}^2+\mathcal{O}_{31}^2<1$.
As in the MSSM framework, these couplings are constrained by the experimental results on Higgs decays, but in the presence of other Higgs states close to the $125$-GeV state, the mixing pattern could be different from that in the MSSM and
$\delta_1$ and/or $\eta_1$ can be sizeable. 
In this case, one may treat $\delta_1$  and $\eta_1$ as free parameters
effectively.
We find that, in this case,
$\mathrm{BR}\left( H_1 \rightarrow \bar{b}s+\bar{s}b\right)$
can be as large as $10^{-4}$.
\end{itemize}

Processes mediated by flavor changing neutral currents (FCNCs)
involving down-type quarks have been largely studied
in the context of 2HDM where significant contributions can be accommodated due to the $\tan\beta$-enhancement of their Yukawa couplings. This type of processes are very useful for investigating the dynamics of quark-flavour mixing, especially the possible non-standard phenomena. Here, our main purpose is studying transitions such as $H_i\rightarrow bs$, keeping always under control other processes that will impose additional experimental constraints, for instance, the B-meson decay $B_s\rightarrow\mu^+\mu^-$ and the mass difference $\Delta M_{B_s}$ \footnote{In the presence of CP phases, limits from electric dipole moments (EDMs) should also be considered. However, in our scenario we have decoupled sfermions, and heavy Higgs masses are above the TeV. In these conditions, the main contributions to EDMs are due to scalar-pseudoscalar Higgs mixing from the two-loop Barr-Zee diagrams with $H_1$. As shown in \cite{Cheung:2014oaa,Inoue:2014nva}, the relevant constraint on our couplings would be, $|\eta_1| \lesssim (0.1 \tan\beta)^2$ which does not playing any relevant role in most of the considered parameter space.}.

\subsection{FC couplings}

It is well-known that, in the MSSM, the superpotential holomorphicity prevents the appearance of Higgs-boson FCNCs by coupling the Higgs-doublet superfield $\Phi_1$ to the down-quark sector and $\Phi_2$ to the up-quark sector. 
However, this property is violated when considering finite radiative (threshold)
corrections due to soft SUSY-breaking interactions~\cite{Banks:1987iu,Hempfling:1993kv,Hall:1993gn,Blazek:1995nv,Carena:1994bv,Pierce:1996zz,Babu:1998er,Borzumati:1999sp,Borzumati:1997bd,Eberl:1999he,Haber:2000kq,Borzumati:2003wb,Borzumati:2003rr}.
As a consequence, Higgs-mediated FCNCs show up at the one-loop level. The general effective Yukawa Lagrangian for down-type quarks may be simply written as \cite{Ellis:2007kb}: 
  \begin{equation}
  -\mathcal{L}^d_{Y}\:=\:\bar{d}^0_R\,{\bf h_d}
{
\left[
 \left({\bf 1}+{\bf \Delta}^{\phi_1}_d\right)\frac{v_1+\phi_1}{\sqrt{2}}
-i\,\left({\bf 1}+{\bf \Delta}^{a_1}_d\right)\frac{a_1}{\sqrt{2}}
\right] }
\,d^0_L\,+\,\bar{d}^0_R\,{\bf h_d}\,
{
\left[
 {\bf \Delta}^{\phi_2}_d\frac{v_2+\phi_2}{\sqrt{2}}
-i\,{\bf \Delta}^{a_2}_d\frac{a_2}{\sqrt{2}}
\right] }
\,d^0_L\:+\:\mathrm{H.c.}
  \label{eq:eqn207}
  \end{equation}
where ${\bf h_d}$ is the tree-level Yukawa matrix and 
$d^0_{L,R}$ refer to the weak eigenstates. After EWSB, this Lagrangian
gives rise to the d-quarks mass terms and also to the Higgs-mediated FC terms. 
For the former, we have:
  \vspace{0.1cm}
  \begin{equation}
  -\mathcal{L}^d_{\mathrm{mass}}\:=\:\frac{v_1}{\sqrt{2}}\,\bar{d}^0_R\,{\bf h_d}\left({\bf 1}+{\bf \Delta}^{\phi_1}_d\right)\,d^0_L\,+\,\frac{v_2}{\sqrt{2}}\,\bar{d}^0_R\,{\bf h_d}\,{\bf \Delta}^{\phi_2}_d\,d^0_L
{+\mathrm{H.c.}}
\:\equiv\:\frac{v_1}{\sqrt{2}}\,\bar{d}^0_R\,{\bf h_d}\,\bigl({\bf 1}
+{\bf \Delta}_d\bigr) \,{d^0_L} 
{+\mathrm{h.c.}}
\label{eq:eqn208}
  \end{equation}
where {\small ${\bf \Delta}_d={\bf \Delta}^{\phi_1}_d+\left(v_2/v_1\right){\bf \Delta}_d^{\phi_2}$ } contains the loop corrections. Transforming the states to the mass basis:
  \begin{eqnarray}
  d^0_R\:=\:{\bf \mathcal{U}}^d_R\,d_R\,;
& \hspace{1.5 cm} & 
d^0_L\:=\:{\bf \mathcal{U}}^d_L\,d_L\:=\:{\bf \mathcal{U}}^Q_L\,{\bf V}\,d_L\,;
\label{eq:eqn209} \\
  u^0_R\:=\:{\bf \mathcal{U}}^u_R\,u_R\,; 
& \hspace{1.5 cm} & 
u^0_L\:=\:{\bf \mathcal{U}}^u_L\,u_L\:=\:{\bf \mathcal{U}}^Q_L\,u_L  \,,
\label{eq:eqn210}
  \end{eqnarray}
we  have:
  \begin{equation}
  -\mathcal{L}^d_{\mathrm{mass}}={\bar{d}_R}\,
\widehat{\bf M}_d\,{d_L} 
{+\mathrm{h.c.}}
\hspace{1.cm} 
\mathrm{with} \quad \widehat{\bf M}_d\,=\,\frac{v_1}{\sqrt{2}}\left({\bf \mathcal{ U}}_R^d\right)^{\dagger}\,{\bf h_d}\,\left[{\bf 1}+{\bf \Delta}_d\right]\,{\bf \mathcal{U}}_L^Q\,{\bf V} \label{eq:eqn211}
  \end{equation}
where
$\widehat{{\bf M}}_d=\mathrm{diag}\left(m_d,m_s,m_b\right)$ is the physical diagonal mass matrix for the down-type quarks. 
Using the flavour basis where ${\bf \mathcal{U}}_L^Q={\bf \mathcal{U}}_R^u={\bf \mathcal{U}}_R^d={\bf 1}$ and introducing ${\bf R}_d={\bf 1}+{\bf \Delta}_d$, we can relate the physical masses to the Yukawa couplings through the following expression:
  \begin{equation}
  {\bf h_d}=\,\frac{\sqrt{2}}{v_1}\,\widehat{{\bf M}}_d\,{\bf V}^{\dagger}\,{\bf R}_d^{-1}\,.
  \label{eq:eqn212}
  \end{equation}
Using this expression in Eq.~(\ref{eq:eqn207}), we obtain the FC effective Lagrangian for the interactions of the physical neutral Higgses with the down-type quarks \cite{Ellis:2007kb,Ellis:2009di}:
  \vspace{0.2cm}
  \begin{equation}
  \mathcal{L}_{\rm FC}\,=\,-\frac{g}{2M_W}\Biggl[ H_i\,\bar{d}\left(\widehat{{\bf
M}}_d\,{\bf g}_{H_i\bar{d}d}^L\,P_L\,+\,{\bf g}_{H_i\bar{d}d}^R\,\widehat{{\bf
M}}_d\,P_R\right)d\Biggr]\,.
  \label{eq:eqn213}
  \end{equation}
%
A simplified expression for the Higgs couplings 
can be obtained working in the single-Higgs-insertion 
(SHI) approximation~\cite{Ellis:2007kb} where:
  \begin{eqnarray}
  & {\bf \Delta}^{\phi_1}_d\,=\,{\bf \Delta}^{a_1}_d\,
=\,{\bf F}_d^0\,; & \nonumber \\
  & {\bf \Delta}^{\phi_2}_d\,=\,{\bf \Delta}^{a_2}_d\,
=\,{\bf G}_d^0\,: & \label{eq:eqn216}
  \end{eqnarray}
  \vspace{-1.cm}
  \begin{eqnarray}
  & {\bf \Delta}_d\,=\,{\bf F}^0_d\,+\,\frac{v_2}{v_1}{\bf G}^0_d\,=\,{\bf F}^0_d\,+\,\tan\beta\,{\bf G}^0_d & \label{eq:eqn217} 
  \end{eqnarray}
For large $\tan\beta$ values, the ${\bf F}_d^0$ term
can be neglected and therefore ${\bf R}_d$ may be approximated as:
  \vspace{-0.2cm}
  \begin{equation}
  {\bf R}_d\,=\,{\bf 1}+\tan\beta\,{\bf G}^0_d\,.
  \label{eq:eqn218}
  \end{equation}
Then the Higgs couplings in the Lagrangian of Eq.~(\ref{eq:eqn213}) will be simplified to:
  \vspace{0.1cm}
  \begin{eqnarray}
  {\bf g}_{H_i\bar{d}d}^L & = & \frac{\mathcal{O}_{1i}}{\cos\beta}{\bf
V}^{\dagger}{\bf R}_d^{-1}{\bf V}\,+\,\frac{\mathcal{O}_{2i}}{\cos\beta}{\bf V}^{\dagger}{\bf R}_d^{-1}{\bf G}_d^0{\bf V}
  \,+\,i\mathcal{O}_{3i}\tan\beta\left[\frac{1}{\sin^2\beta}{\bf V}^{\dagger}{\bf R}_d^{-1}{\bf V}\,-\,\frac{1}{\tan^2\beta}
\right]\,, \label{eq:eqn214} \\
  & & \nonumber \\
  {\bf g}_{H_i\bar{d}d}^R & = & \left({\bf g}_{H_i\bar{d}d}^L\right)^{\dagger}\,.
 \label{equation215}
  \end{eqnarray}
By noting {\small ${\bf V}^{\dagger}{\bf R}_d^{-1}{\bf G}_d^0{\bf V}=\left({\bf 1}-{\bf V}^{\dagger}{\bf R}_d^{-1}{\bf V}\right)/\tan\beta$}, we observe that the size of flavour violation is dictated by the off-diagonal components of the matrix {\small ${\bf V}^{\dagger}{\bf R}_d^{-1}{\bf V}={\bf V}^{\dagger}\left({\bf 1}+\tan\beta~{\bf G}^0_d\right)^{-1}{\bf V}$}. Therefore the Higgs couplings to the down-type quarks will be determined once {\small ${\bf G}_d^0$} is known. 

The detailed expression for this quantity {\small ${\bf G}_d^0$} in
terms of soft SUSY-breaking parameters
can be found in Appendix~\ref{sec:appendixa}. 
Observe that in this formalism, 
flavour violation from off-diagonal components of the sfermion mass matrices 
as well as its diagonal parts has been included.

\subsection{FC Higgs decays}

For the computation of  
the decay width $\Gamma(H_i\rightarrow\,\bar{b}s+\bar{s}b)$,
we consider the relevant terms involving $b$ and $s$ quarks 
in Eq.~(\ref{eq:eqn213}). 
Introducing the effective couplings
$y_{L_i}\equiv m_b{\bf g}_{H_i\bar{b}s}^L/v$ and $y_{R_i}\equiv m_s{\bf g}_{H_i\bar{b}s}^R/v$, we may write:
  \begin{eqnarray}
  \mathcal{L}_{H_ibs} & = & -H_i\,\bar{b}\Bigl[y_{L_i}\,P_L+y_{R_i}\,P_R\Bigr]\,s\:+\:\mathrm{H.c.}
  \label{eq:eqn226}
  \end{eqnarray}
This expression can be rewritten as:
  \begin{eqnarray}
  \mathcal{L}_{H_ibs} & = & -H_i\,\bar{b}\Bigl[g^S_i\,+\,ig^P_i\gamma_5\Bigr]\,s\:+\:{\rm H.c.}     
  \label{eq:eqn2262}
  \end{eqnarray}
with {$g^S_i=(y_{L_i}+y_{R_i})/2$ and $g^P_i=i(y_{L_i}-y_{R_i})/2$.}
Then, using Eq.~(28) of \cite{Lee:2003nta}, the decay width can be obtained as:
  \begin{eqnarray}
  \Gamma\left(H_i\rightarrow\,\bar{b}s+\bar{s}b\right) & = & 2\times\frac{N_c\,m_{H_i}}{16\pi}\lambda^{1/2}\left(1,x_b,x_s\right)\,\kappa_{QCD}\times \nonumber \\
  & & \left[\left(1-x_b-x_s\right)\left(\left|y_{L_i}\right|^2+\left|y_{R_i}\right|^2\right)-4\sqrt{x_b x_s}\,\Re e\left(y_{L_i}y^*_{R_i}\right)\right]
  \label{eq:eqn227}
  \end{eqnarray}
where $\kappa_{QCD}=1+5.67 \,\alpha_S\left(m^2_{H_i}\right)/\pi$ including the QCD correction, $x_f=m^2_f/m^2_{H_i}$ and $\lambda\left(1,a,b\right)=(1-a-b)^2-4ab$. Note that in $y_{L_i,R_i}$ the masses involved are {\small $m_{b,s}=m_{b,s}\left(m_{H_i}\right)$}.
\section{Experimental Constraints} \label{sec:exconstr}
In our analyses, we use two main sets of experimental data: collider constraints on Higgs and supersymmetric particles and constraints on flavour changing processes. 
Collider constraints come mainly from CMS and ATLAS, 
the two general purpose experiments in the LHC 
that claimed the observation of a new $125$ 
GeV particle in 2012 \cite{Aad:2012tfa,Chatrchyan:2012ufa}. 
Regarding indirect processes, we use the current flavour experimental 
data associated with B-meson decays and mass differences. 
\subsection{Collider Constraints} \label{subsec:colliders}
ATLAS and CMS are the two general purpose LHC experiments which provide the most accurate data concerning the Higgs boson and SUSY. In particular, we will consider the Higgs $\gamma\gamma$ signal, the $\tau\tau$-channel limits, and direct limits on supersymmetric particle masses. All these constraints have been already used in previous works \cite{Barenboim:2013bla, Barenboim:2013qya}, so we refer to them for details. Here, we will summarize the basic requirements taken into account during the analysis. First, according to the experimental data so far, we require a diphoton signal in the range:
  \vspace{-0.35cm}
  \begin{equation}
  0.75\,\leq\,\mu_{\gamma\gamma}^{LHC}\,\leq\,1.55\, ,
  \label{eq:4102}
  \end{equation}
where $\mu_{X}$ is the signal strength for a Higgs decaying to $X$: 
$\mu_{X}=[\sigma(pp\to H)\times\mbox{BR}(H\to X)]/
[\sigma(pp\to H)_{\rm{SM}}\times\mbox{BR}(H\to X)_{\rm{SM}}]$.

On the other hand, we apply the limits set by CMS \cite{CMS:2013hja} and 
ATLAS \cite{Aad:2014vgg} in the $H \to \tau\tau$-channel. 
Specifically, we use the 95 \% Confidence Level (CL) limits
on the gluon-fusion and $b$-associated Higgs boson production cross sections
times the branching ratio into $\tau$ pairs,
presented in
Fig.~4 in Ref.~\cite{CMS:2013hja} and Fig.~11 in Ref.~\cite{Aad:2014vgg}.
In these analysis, extended searches for extra Higgs states have been carried out 
for masses up to $1$ TeV at $95\%$ CL. In our case, these limits will be imposed 
to all three neutral Higgs states: $H_1$, $H_2$ and $H_3$.

Finally, we must take into account direct bounds on SUSY masses. Taking into account that the effects of SUSY particles on Higgs couplings are non-decoupling, we can apply conservative limits on the masses. For the gluino, we set the mass limit at {$m_{\tilde{g}}\gtrsim 1.4$} TeV when the neutralino mass is below $\sim 700$ GeV, in agreement with the exclusion limits from ATLAS \cite{ATLAS:2014agr} and CMS \cite{CMS:2014cms}. The mass limits for the third generation squarks are taken from ATLAS data \cite{ATLAS:2014agr} as: $m_{\tilde{t}_1}\,\gtrsim\,650$ GeV when the neutralino mass is below $250$ GeV, or $m_{\tilde{t}_1}-m_{\chi_1^0}\lesssim 175$ GeV when it is nearly degenerate with the LSP. Finally, according to ATLAS searches \cite{ATLAS:2014agr}, the chargin mass limits are: {$m_{\tilde{\chi}^{\pm}_1} \gtrsim 700$} GeV for dominant decays into charged leptons, or {$m_{\chi^{\pm}_1}\gtrsim 450$} GeV when the decays into weak bosons prevail.

\subsection{FC Constraints} \label{subsec:fc}

Apart from these data coming from the collider experiments, there is another kind of processes that can play a significant part in the search for SM extensions. As presented in the previous works 
\cite{Barenboim:2013bla,Barenboim:2013qya}, 
flavour constraints may be a powerful weapon to restrict the parameter space, 
especially the parameter $\tan\beta$, 
even in the absence of complex flavour structures beyond the SM Yukawa couplings. Therefore, for our analysis, we will make use of indirect bounds coming from B-meson decays and mass differences. In particular, we will consider: $B^0_s\rightarrow\mu^+\mu^-$, $\Delta M_{B_s}$ and $B\rightarrow X_s\gamma$. In the case of the rare decay $B^0_s\rightarrow\mu^+\mu^-$, the latest experimental value for its branching fraction is the combined analysis of CMS and LHCb data at $7$ TeV and $8$ TeV, with integrated luminosities of $25$ $\mathrm{fb}^{-1}$ and $3$ $\mathrm{fb}^{-1}$ respectively \cite{CMS:2014xfa}:
  \begin{equation}
  \mathrm{BR}\left(B^0_s\rightarrow\mu^+\mu^-\right)\,=\,\left(2.8^{+0.7}_{-0.6}\right)\cdot 10^{-9}
  \end{equation}
Hence, our analysis bound at $2\sigma$ will be:
  \begin{equation}
  \mathrm{BR}\left(B^0_s\rightarrow\mu^+\mu^-\right)\,\leq\,4.2\cdot 10^{-9}
  \end{equation}
  
Another valuable flavour decay is $B\rightarrow X_s\gamma$, which becomes the most restrictive constraint for medium and low $\tan\beta$ values. Combining the BaBar, Belle and CLEO analysis, the world average value given by HFAG \cite{Amhis:2014hma} is:
  \begin{equation}
  \mathrm{BR}\left(B\rightarrow X_s\gamma\right)\,=\,\left(3.43\pm 0.21\pm 0.07\right)\cdot 10^{-4}
  \end{equation}

For the opposite $\tan\beta$ regime, that is large $\tan\beta$ values, the main experimental result turns out to be the $\mathrm{B_s}$-meson mass difference $\Delta M_{B_s}$. The present experimental value is \cite{Amhis:2014hma}:
  \begin{equation}
  \Delta M_{B_s}\,=\,\left(17.757\,\pm\,0.021\right)\quad\mathrm{ps}^{-1}
  \end{equation}
We will require in our analysis:
  \begin{equation}
  15.94\,\mathrm{ ps}^{-1}\:\leq\:\Delta M_{B_s}\:\leq\:19.83\,\mathrm{ps}^{-1}
  \end{equation}
where we included the theoretical error on $f_{B_s} \sqrt{B_s} = 262 \pm 10$ \cite{Aoki:2013ldr}.

\section{Numerical Analysis} \label{sec:numanalysis}
Before we present the results of our numerical analysis, we present first some approximate
analytic expressions for ${\bf G}_d^0$, ${\bf R}_d^{-1}$ and, finally,
$\left({\bf V}^\dagger {\bf R}_d^{-1} {\bf V}\right)_{32,23}$, 
which are most relevant to the FC Higgs 
decay into $b$ and $s$ quarks. Note that the FC structure is common to the two
situations under consideration: the full MSSM framework and the generic Supersymmetric SM.

From Eq.~(\ref{eq:eqn214}) and using 
${\bf V}^{\dagger}{\bf R}_d^{-1}{\bf G}_d^0{\bf V}=\left({\bf 1}-{\bf
V}^{\dagger}{\bf R}_d^{-1}{\bf V}\right)/\tan\beta$, 
we observe that
the size of flavour violation is dictated by the off-diagonal components of 
the matrix ${\bf V}^{\dagger}{\bf R}_d^{-1}{\bf V}$ which can be determined
once ${\bf G}_d^0$ is known. 

As shown in Appendix~\ref{sec:appendixa},  one may need 
the explicit forms of the flavour violating matrices 
${\bf \delta}\widetilde{{\bf M}}^2_{Q,U,D}$ and $\mathbf{\delta a}_u$
to derive ${\bf G}_d^0$. 
Assuming universality for the first two generations, we introduce the following
flavour parametrization \footnote{We are assuming, for simplicity, symmetric Yukawa
and trilinear matrices at tree-level. In general, $\left({\bf h}_u^{-1}{\bf
a}_u\right)_{23}$ and  $\left({\bf h}_u^{-1}{\bf a}_u\right)_{32}$ can be different
from each other and complex.}
  \begin{equation}
  \widetilde{\bf M}^2_X=\left(\begin{array}{ccc}
  \rho & 0 & 0\\
  0 & \rho & \delta_X\\
  0 & \delta_X & 1
  \end{array}\right)\,\widetilde{M}^2_{X_3}\,,
  \hspace{1cm}
  {\bf h}_u^{-1}{\bf a}_u=\left(\begin{array}{ccc}
  \rho & 0 & 0\\
  0 & \rho & \delta_{A_u} \\
  0 & \delta_{A_u} & 1
  \end{array}\right)\, A_{u_3}
  \label{eq:eqn502}
  \end{equation}
  \vspace{0.1cm}
where $X=Q,U,D$ and  then ${\bf \delta}\widetilde{\bf M}_{Q,U,D}^2$ and
${\bf \delta a}_{u}$ are given by
  \begin{equation}
  {\bf \delta}\widetilde{\bf M}_{Q,U,D}^2\,=\,\widetilde{\bf
M}_{Q,U,D}^2\,-\,\widetilde{M}_{Q,U,D}^2{\bf 1} \,,
\hspace{2.cm} 
{\bf \delta a}_{u}\,=\,{\bf a}_{u}\,-\,{\bf h}_{u}A_{u}\,
  \label{eq:eqn224}
  \end{equation}
with $\widetilde{M}_{Q,U,D}^2=\frac{1}{3}\mathrm{Tr}\left(\widetilde{\bf
M}_{Q,U,D}^2\right)=\frac{1}{3}\left(2\rho
+1\right)\,\widetilde{M}^2_{{Q_3,U_3,D_3}}$ and
$A_{u}=\frac{1}{3}\mathrm{Tr}\left({\bf h}_{u}^{-1}{\bf
a}_{u}\right)=\frac{1}{3}\left(2\rho+1\right)\,A_{u_3}$.
In the basis where the up-type Yukawa quarks are diagonal ${\bf
h}_u=\mathrm{diag}\left(y_u,y_c,y_t\right)$, we have,
  \begin{equation}
 {\bf \delta}\widetilde{{\bf M}}^2_X=\left(\begin{array}{ccc}
  \frac{\rho-1}{3} & 0 & 0\\
  0 & \frac{\rho-1}{3} & \delta_X\\
  0 & \delta_X & -\frac{2}{3}\left(\rho-1\right)
\end{array}\right)\,\widetilde{M}^2_{X_3}\,,
  \hspace{1cm}
  \mathbf{\delta a}_u=\left(\begin{array}{ccc}
  \frac{\rho-1}{3}\,y_u & 0 & 0\\
  0 & \frac{\rho-1}{3}\,y_c & \delta_{A_u}\,y_c\\
  0 & \delta_{A_u}\,y_t & -\frac{2}{3}\left(\rho-1\right)\,y_t
  \end{array}\right)\, A_{u_3}\,.
  \label{eq:eqn503}
  \end{equation}
%
%
Inserting the above expression for ${\bf \delta}\widetilde{\bf M}_{Q,U,D}^2$ and 
${\bf \delta a}_{u}$  into Eqs.~(\ref{eq:eqn220})
and (\ref{eq:eqn221}), we obtain 
\begin{equation}
{\bf G}^0_d\simeq\left(\begin{array}{ccc}
      \epsilon & 0 & 0\\
      0 & \epsilon & \delta\epsilon_{23}\\
        0 & \delta\epsilon_{32} & \epsilon+\eta \end{array}\right)
\label{eq:eqn504}
\end{equation}
%
where $\epsilon$, $\eta$, $\delta\epsilon_{32}$ and $\delta\epsilon_{23}$ are parameters
containing the main loop contributions of Eq.~(\ref{eq:eqn220}) and
Eq.~(\ref{eq:eqn221}). Here, it is worth mentioning that EW corrections in those
two equations have been neglected, while the down-type Yukawa couplings have
been approximated as ${\bf h}_d\simeq\frac{\sqrt{2}}{v_1}\widehat{{\bf M}}_d{\bf
V}^{\dagger}$. 
The explicit forms of the diagonal entries $\epsilon$ and $\eta$ are given in
Appendix~\ref{sec:appendixa},
while the off-diagonal elements, which are the key ones for us as will be
seen later, are given by the following expressions:
{\small
  \begin{eqnarray}
  \delta\epsilon_{23} & = & \delta_L \left[\frac{2\alpha_s}{3\pi}\, \mu^*\,
M_3^*\,\tilde{M}^2_{Q_3}\,K\left(\tilde{M}^2_Q, \tilde{M}^2_D, \left| M_3 \right|^2
\right)+\,\frac{\left|y_t\right|^2}{16\pi^2}\,\mu^*\,A_t^*\,\tilde{M}^2_{Q_3}\,K\left(\tilde{M}^2_Q,
\tilde{M}^2_D, \left| \mu \right|^2 \right)\right] \label{eq:eqn507} \\
       &   & \hspace{-0.35cm} +\,\delta_{A_u}
\left[\frac{\left|y_t\right|^2}{16\,\pi^2}\mu^* A^*_t I\left(\tilde{M}^2_Q,
\tilde{M}^2_U, \left| \mu \right|^2\right)\right]\,+\,\delta_R
\left[\frac{2\alpha_s}{3\pi}\,\frac{V^{^*}_{33}\,y_b}{V^{^*}_{22}\,y_s} \mu^*\,
M_3^*\, \tilde{M}^2_{D_3} K\left(\tilde{M}^2_D, \tilde{M}^2_Q, \left| M_3
\right|^2\right)\right] \nonumber \\
       &   &
\hspace{-0.35cm}+\,\left(\rho-1\right)\left[\frac{2\alpha_s}{3\pi}\,\frac{V^{^*}_{32}}{V^{^*}_{22}}\mu^*\,
M_3^*\, \tilde{M}^2_{D_3} K\left(\tilde{M}^2_D, \tilde{M}^2_Q, \left| M_3
\right|^2\right)\right] \nonumber \\
       &   &  \nonumber \\
  \delta\epsilon_{32} & = & \delta_L\left[\frac{2\alpha_s}{3\pi}\, \mu^*\,
M_3^*\,\tilde{M}^2_{Q_3}\,K\left(\tilde{M}^2_Q, \tilde{M}^2_D, \left| M_3 \right|^2
\right)\right] 
\,+\, \delta_{A_u}\left[\frac{\left|y_c\right|^2}{16\pi^2}
\mu^*\,A^*_t\,I\left(\tilde{M}^2_Q, \tilde{M}^2_U, \left| \mu
\right|^2\right)\right] \label{eq:eqn508} \\
      &    &
\hspace{-0.52cm}+\,
\delta_R\left[\frac{2\alpha_s}{3\pi}\left(\frac{V^{^*}_{22}\,y_s}{V^{^*}_{33}\,y_b}\,-\,\frac{V^{*^2}_{23}\,y_b}{V^{^*}_{22}\,V^{^*}_{33}\,y_s}\right)\mu^*\,M^{^*}_3\tilde{M}^2_{D_3}\,K\left(\tilde{M}^2_D,
\tilde{M}^2_Q, \left| M_3 \right|^2 \right) \right] \nonumber \\
& &
\hspace{-0.52cm}-\,
\left(\rho -1\right)
\left[\frac{2\alpha_s}{3\pi}\frac{V^{^*}_{23}}{V^{^*}_{33}}\, \mu^*\,
M_3^*\,\tilde{M}^2_{D_3}\,K\left(\tilde{M}^2_D, \tilde{M}^2_Q, \left| M_3 \right|^2
\right)\right] \nonumber
  \end{eqnarray}}
where $\delta_Q \equiv \delta_{L}$ and $\delta_U=\delta_D\equiv\delta_{R}$. 
We note that there are four types of flavour-violating terms proportional to
$\delta_L$, $\delta_{A_u}$, $\delta_R$, and $(\rho-1)$ \footnote{ Please note that our definition of $\delta_{A_u}$ in Eq~(\ref{eq:eqn503}) makes it different from the $\delta_{LR}$ usually defined in the literature.}.
Then, we obtain
\begin{equation}
  {\bf R}^{-1}_d \simeq \left(\begin{array}{ccc}
           \frac{\left(1+\epsilon\tan\beta\right)\left(1+(\epsilon+\eta)\tan\beta\right)-\delta\epsilon_{23}\delta\epsilon_{32}\tan^2\beta}{\mathrm{Det}\left({\bf
R}_d\right)} & 0 & 0\\
           0 &
\frac{\left(1+\epsilon\tan\beta\right)\left(1+(\epsilon+\eta)\tan\beta\right)}{\mathrm{Det}\left({\bf
R}_d\right)} &
-\frac{\delta\epsilon_{23}\left(1+\epsilon\tan\beta\right)\tan\beta}{\mathrm{Det}\left({\bf
R}_d\right)}\\
           0 &
-\frac{\delta\epsilon_{32}\left(1+\epsilon\tan\beta\right)\tan\beta}{\mathrm{Det}\left({\bf
R}_d\right)} & \frac{\left(1+\epsilon\tan\beta\right)^2}{\mathrm{Det}\left({\bf
R}_d\right)}\end{array}\right)
  \label{eq:eqn510}
  \end{equation}
from 
  \begin{equation}
  {\bf R}_d\:=\:{\bf 1}+\tan\beta\, {\bf G}^0_d\:\simeq \:\left(\begin{array}{ccc}
        1+\epsilon\tan\beta & 0 & 0\\
        0 & 1+\epsilon\tan\beta & \delta\epsilon_{23}\tan\beta\\
        0 & \delta\epsilon_{32}\tan\beta & 1+\left(\epsilon+\eta\right)\tan\beta
        \end{array}\right)\,.
  \label{eq:eqn509}
  \end{equation}
\vspace{0.5cm}
As we will see, the most relevant flavour-violating matrix element
in the FC Higgs decay $H_i\to b s$
is $\left({\bf V^{\dagger} R^{-1}_d V}\right)_{32,23}$.
In principle, this matrix element $\left({\bf V^{\dagger} R^{-1}_d V}\right)_{32}$
is:
  \begin{eqnarray}
  \left({\bf V}^{\dagger} {\bf R}^{-1}_d {\bf V}\right)_{32} & = & \sum_{i,j=1}^{3}\,
{\bf V}^*_{i3}\, \left({\bf R}^{-1}_{d}\right)_{ij}\, {\bf V}_{j2} \nonumber \\
  & = & \sum_{i=1}^{3}\left[ {\bf V}^*_{i3}\, \left({\bf R}^{-1}_{d}\right)_{ii}\,
{\bf V}_{i2}\right] + {\bf V}^*_{23}\, \left({\bf R}^{-1}_{d}\right)_{23}\, {\bf V}_{32} + {\bf V}^*_{33}\,
\left({\bf R}^{-1}_{d}\right)_{32}\, {\bf V}_{22}\,.
  \label{eq:eqn521}
  \end{eqnarray}
  \vspace{0.1cm}
Using Eq.~(\ref{eq:eqn510}) for ${\bf R}^{-1}_d$ and taking into account 
the unitarity of the CKM matrix, we can write:
  \begin{equation}
  \sum_{i=1}^{3}\left[ {\bf V}^*_{i3}\, \left({\bf R}^{-1}_{d}\right)_{ii}\, 
{\bf V}_{i2}\right] =
\left({\bf V}^*_{13}{\bf V}_{12}+{\bf V}^*_{23}{\bf V}_{22}\right)\frac{\left(1+\epsilon\tan\beta\right)\eta\,\tan\beta}{\mathrm{Det}\left({\bf R}_d\right)}
-{\bf V}^*_{13}{\bf V}_{12}\frac{\delta\epsilon_{23}\delta\epsilon_{32}\tan^2\beta}{\mathrm{Det}\left({\bf R}_d\right)}\,.
  \label{eq:eqn522} 
  \end{equation}
  \vspace{0.1cm}
In this expression, we can neglect all the terms proportional to 
${\bf V}^*_{13}{\bf V}_{12}\sim 8 \times 10^{-4}$ 
with respect to  ${\bf V}^*_{23}{\bf V}_{22}\sim 4 \times 10^{-2}$, even for
the last term proportional to $\delta\epsilon_{23}\delta\epsilon_{32}$, which, as
can be seen from Eqs.~(\ref{eq:eqn507},\ref{eq:eqn508}) and (\ref{eq:eqn505},\ref{eq:eqn506}), is, 
for sizeable mass insertion (MI), 
of the same order as $\epsilon \times \eta$.

Then, in Eq.~(\ref{eq:eqn521}), if we have similar values of the off-diagonal
elements $\left({\bf R}^{-1}_d\right)_{23}$ and 
$\left({\bf R}^{-1}_d\right)_{32}$, the former
can also be
neglected with respect to later, being suppressed by an additional
$|{\bf V}^*_{23}{\bf V}_{32}|\sim 2.10^{-3}$. 
Therefore, in the presence of sizeable mass insertions $\delta_{L,R} \geq {\bf V}^*_{23}{\bf V}_{22}$, we have $\delta \epsilon_{32} \geq  \eta \times {\bf V}^*_{23}{\bf V}_{22}$ and then we can safely take,
  \begin{eqnarray}
  \left({\bf V^{\dagger} R^{-1}_d V}\right)_{32}
& \simeq & {\bf V}^*_{33}\, {\bf V}_{22}\, \left({\bf R^{-1}_{d}}\right)_{32} \,+ \,{\bf V}^*_{23}{\bf V}_{22} \frac{\left(1+\epsilon\tan\beta\right)\eta\,\tan\beta}{\mathrm{Det}\left({\bf R}_d\right)}
\simeq \left({\bf R^{-1}_{d}}\right)_{32}\,.
  \label{eq:eqn523}
  \end{eqnarray}
Repeating the same exercise with  $\left({\bf V^{\dagger} R^{-1}_d V}\right)_{23}$, we obtain,
  \begin{eqnarray}
  \left({\bf V^{\dagger} R^{-1}_d V}\right)_{23}
& \simeq & {\bf V}^*_{22}\, {\bf V}_{33}\, \left({\bf R^{-1}_{d}}\right)_{23} \,+ \,{\bf V}^*_{22}{\bf V}_{23} \frac{\left(1+\epsilon\tan\beta\right)\eta\,\tan\beta}{\mathrm{Det}\left({\bf R}_d\right)}
\simeq \left({\bf R^{-1}_{d}}\right)_{23}\,.
  \label{eq:eqn523b}
  \end{eqnarray}
The study of these matrix elements is very interesting because of their dependence on
$\delta_{L}$, $\delta_{R}$ and $\delta_{A_u}$. Looking at Eq.~(\ref{eq:eqn508}) for
$\delta\epsilon_{32}$ and comparing the $\delta_{L}$ and $\delta_{R}$ contributions,
we can see that the $\delta_{R}$ term is suppressed by the difference
$\left(y_s/y_b - {{\bf V}_{23}^*}^2 y_b/y_s\right) \simeq - 0.013$ 
with respect to the
$\delta_L$ term. Therefore, these matrix elements, and especially {\small
$\left({\bf R}^{-1}_{d}\right)_{32}$}, will have very different values for these two mass
insertions. Also from these equations, it is evident that, if
 $\delta_{L}\simeq \delta_{A_u}$, we would obtain similar results from both types of mass insertions to $\delta \epsilon_{23}$, but its contributions to $\delta \epsilon_{32}$ would be suppressed by an additional factor $(y_c/y_t)^2$. In any case, the $\delta_{A_u}$ contributions to $\delta \epsilon_{23}$ and $\delta \epsilon_{32}$ are always smaller than the $\delta_{R}$ contributions. For this reason, we will only consider the cases of the $\delta_{L}$ and $\delta_{R}$ insertions

All the results of the numerical analysis presented in the following sections and the corresponding figures are done with the 
{\tt CPsuperH2.3} code \cite{Lee:2003nta,Lee:2007gn,Lee:2012wa}.

\subsection { Full MSSM framework}

In the full MSSM framework, the effective FC Higgs couplings are given by: 
  \vspace{0.05cm}
  \begin{eqnarray}
  y_{L_i} & \equiv & \frac{m_b}{v}{\bf
g}_{H_i\bar{b}s}^L\,=\,
\frac{m_b}{v}\Biggl(\frac{\mathcal{O}_{1 i}}{\cos \beta}\,-\,
\frac{\mathcal{O}_{2 i}}{\sin\beta}\,+\,i\tan \beta\frac{\mathcal{O}_{3 i}}
{\sin^2\beta}\Biggr)\left({\bf V}^\dagger{\bf R}_d^{-1}{\bf V}\right)_{32} \,,
\label{eq:eqn511} \\
  y_{R_i} & \equiv &
\frac{m_s}{v}{\bf g}_{H_i\bar{b}s}^R\,=\,
\frac{m_s}{v}\Biggl(\frac{\mathcal{O}_{1 i}}{\cos \beta}\,-\,
\frac{\mathcal{O}_{2 i}}{\sin\beta}\,-\,i\tan \beta\frac{\mathcal{O}_{3 i}}
{\sin^2\beta}\Biggr)\left({\bf V}^\dagger{\bf R}_d^{-1}{\bf V}\right)_{23}^*\,,
\label{eq:eqn512}
  \end{eqnarray}
  \vspace{0.05cm}
and the decay width, Eq.~(\ref{eq:eqn227}), is:
  \begin{eqnarray}
  \Gamma\left(H_i\rightarrow\bar{b}s+\bar{s}b\right) & \simeq & \frac{3 m_{H_i}}{8 \pi}\,\kappa_{QCD}\,\left(|y_{L_i}|^2+|y_{R_i}|^2 \right) 
\label{eq:eqn513} \\
  & \simeq & \frac{3 m_{H_i}m_b^2}{8\pi
v^2}\,\kappa_{QCD}\,\left[\left(\frac{\mathcal{O}_{1i}}{\cos\beta}-\frac{\mathcal{O}_{2i}}{\sin\beta}\right)^2+\left(\tan\beta\frac{\mathcal{O}_{3i}}{\sin^2\beta}\right)^2\right] \nonumber \\
&& \times 
\left(\Bigl|\left({\bf V}^\dagger{\bf R}_d^{-1}{\bf V}\right)_{32}\Bigr|^2 
\,+\, \frac{m_s^2}{m_b^2} \Bigl|\left({\bf V}^\dagger{\bf R}_d^{-1}{\bf V}\right)_{23}\Bigr|^2 \right)\,, \nonumber 
  \end{eqnarray}
  \vspace{0.05cm}  
%
In the case of $\delta_{L}$ insertions, from the discussion in
the previous section, we observe $\left|y_{R_i}\right|\ll\left|y_{L_i}\right|$
due to the $m_s/m_b$ suppression with 
$\left({\bf V}^\dagger{\bf R}_d^{-1}{\bf V}\right)_{32} \sim 
 \left({\bf V}^\dagger{\bf R}_d^{-1}{\bf V}\right)_{23}^*$.
In the presence of $\delta_{R}$ insertions the situation is more involved and both terms must be considered.

Now, considering the total decay widths of the Higgs bosons, 
we will obtain the corresponding branching ratios. 
In the case of the lightest Higgs, its total decay width is dominated by 
the decay into two $b$-quarks, two $W$-bosons and two $\tau$-leptons. Thus:
  \begin{eqnarray}
  \Gamma_{H_1} & = & \frac{m_{H_1}m_b^2}{8\pi v^2}\Biggl[\left(3\,\kappa_{QCD}
+\frac{m_\tau^2}{m_b^2}\right) \tan^2\beta\,\left(\mathcal{O}_{11}^2 +
\mathcal{O}_{31}^2\right)\,+\,I_{PS}\frac{m_{H_1}^2}{m_b^2}\left(\mathcal{O}_{21}+\frac{\mathcal{O}_{11}}{\tan \beta}\right)^2\Biggr]
  \label{eq:eqn514}
  \end{eqnarray}
  \vspace{0.1cm}
where in the large-$\tan\beta$ limit we have used
$\left(\mathcal{O}_{11}^2/\cos^2\beta + \tan^2\beta\,\mathcal{O}_{31}^2
\right)\simeq \tan^2\beta\left(\mathcal{O}_{11}^2 + \mathcal{O}_{31}^2\right)$
and $\left(\sin\beta\,\mathcal{O}_{21}+\cos\beta\,
\mathcal{O}_{11}\right)^2\simeq \left(\mathcal{O}_{21}+\mathcal{O}_{11}/\tan \beta\right)^2$. $I_{PS}$ in the second term refers to the phase-space integral in the Higgs decay into two W bosons \cite{Lee:2003nta} and can be approximated by $I_{PS}\simeq6.7\times10^{-4}$ when $m_{H_1}=125$ GeV. Then, the branching ratio will be:
  {\small   
  \begin{equation}
  \mathrm{BR}\left( H_1 \rightarrow \bar{b}s+\bar{s}b\right)\,=\,\Bigl| \left(
{\bf V}^{\dagger} {\bf R}^{-1}_d {\bf V}\right)_{32} \Bigr|^2
\frac{3\,\kappa_{QCD}\left[\left(\frac{\mathcal{O}_{11}}{\cos\beta}-\frac{\mathcal{O}_{21}}{\sin\beta}\right)^2+\left(
\frac{\mathcal{O}_{31}}{\sin\beta\cos\beta}\right)^2\right]}{\left(3\,\kappa_{QCD}+\frac{m^2_{\tau}}{m_b^2}\right)\tan^2\beta\left(\mathcal{O}_{11}^2+\mathcal{O}_{31}^2
\right)+I_{PS}\frac{m^2_{H_1}}{m_b^2}\left(\mathcal{O}_{21}+\frac{\mathcal{O}_{11}}{\tan\beta}\right)^2}\,.
  \label{eq:eqn515}  
  \end{equation}}
\vspace{0.1cm} 
For the heavier Higgses, and for $\tan\beta\gtrsim 30$, the total decay width is dominated by the bottom and tau widths: 
  \begin{equation}
  \Gamma_{H_2}\,\simeq\,\frac{m_{H_2}m_b^2}{8\pi
v^2}\left(3\,\kappa_{QCD}\,+\,\frac{m_\tau^2}{m_b^2}\right)\,\tan^2\,\beta\left(\mathcal{O}_{12}^2
+ \mathcal{O}_{32}^2\right)
  \label{eq:eqn516}
  \end{equation}
  \vspace{0.1cm}
using 
$\left(\mathcal{O}_{12}^2/\cos^2\beta + \tan^2\beta\,\mathcal{O}_{32}^2 \right)
\simeq \tan^2\beta\left(\mathcal{O}_{12}^2 + \mathcal{O}_{32}^2\right)$
and the branching ratio is:
  \begin{equation}
  {\rm
BR}\left(H_2\to\bar{b}s+\bar{s}b\right)\,=\,3\,\kappa_{QCD}\left|\left({\bf
V}^{\dagger}{\bf R}_d^{-1}{\bf
V}\right)_{32}\right|^2\frac{\quad\left(\frac{\mathcal{O}_{12}}{\cos\beta}-\frac{\mathcal{O}_{22}}{\sin\beta}\right)^2\,+\,\left(\frac{\mathcal{O}_{32}}{\sin\beta\cos\beta}\right)^2\quad}{\left(3\,\kappa_{QCD}\,+\,\frac{m_\tau^2}{m_b^2}\right)\tan^2\beta\left(\mathcal{O}_{12}^2
+ \mathcal{O}_{32}^2\right)}\,.
  \label{eq:eqn517}
  \end{equation}
  \vspace{0.1cm}  
In a previous work \cite{Barenboim:2013qya}, we showed that the latest LHC data for the Higgs signal in the diphoton channel strongly constrains the Higgs mixing within a general CP-violating model. In particular, if we consider the lightest Higgs as the recently discovered boson at $126$ GeV, its mixing conditions are$\left({\cal O}^2_{11}+{\cal O}^2_{31}\right)\sim 1/\tan^2\beta$
and ${\cal O}^2_{21}\sim 1$. Additionally, using the parametrization presented in Eq.~(\ref{eq:eqn206}), we have:
  \begin{eqnarray}
  \mathrm{BR}\left(H_1 \rightarrow \bar{b}s+\bar{s}b\right) & = & \Bigl|
\left({\bf V^{\dagger} R^{-1}_d V}\right)_{32} \Bigr|^2 \frac{3\,\kappa_{QCD}\,(\delta_1^2 + \eta_1^2)}{\quad\left(3\,\kappa_{QCD}+\frac{m^2_{\tau}}{m_b^2}\right)+I_{PS}\frac{m^2_{H_1}}{m_b^2}\quad}\,. \label{eq:eqn519}
  \end{eqnarray}
Still following \cite{Barenboim:2013qya}, it was also showed that the diphoton
condition establishes for the heavier Higgs mixings that: {\small
$\left({\cal O}^2_{1i}+{\cal O}^2_{3i}\right)\sim 1$} and {\small $\mathcal{O}_{2i}\lesssim 1/\tan\beta$}, {\small $i=2,3$}. Hence:
  \begin{eqnarray}
  \mathrm{BR}\left(H_2\rightarrow \bar{b}s+\bar{s}b\right) & \simeq & \left| 
\left({\bf V^{\dagger} R^{-1}_d V}\right)_{32} \right|^2 \frac{3\,\kappa_{QCD}}{\quad 3\,\kappa_{QCD}+\frac{m^2_{\tau}}{m_b^2}\quad } \label{eq:eqn520}
  \end{eqnarray}
where we have considered $\tan\beta \geq 3$ and, therefore, $1/\cos\beta\simeq\tan\beta$ and
$\sin\beta\simeq1$ in a good approximation. 

\subsubsection { Left-handed (L) insertion}
First, we analyze the case with $\delta_{L}\neq0$ and $\delta_{R}=0$:
  \begin{eqnarray}
  \delta\epsilon_{32(L)} & \simeq & \delta_{L}\,\frac{2\alpha_s}{3\pi}\,\mu^*\, M_3^*\,  \tilde{M}^2_{Q_3} K\left(\tilde{M}^2_Q, \tilde{M}^2_D, \left| M_3 \right|^2 \right) \sim -3\cdot10^{-3} \delta_{L}
  \label{eq:eqn524}
  \end{eqnarray}
where we can see that $\delta\epsilon_{32(L)}$ has a non-decoupling behaviour 
as it depends only on ratios of sparticle masses and 
we have used $\tilde{M}_{Q,D,Q_3}\sim 5$ TeV, $M_3\sim 7$ TeV and $\mu\sim 6$ TeV.
Consequently, for the maximum values of $\tan\beta$ and $\delta_{L}$ 
considered during the scan, $\delta_{L} \sim 0.5$ \footnote{Notice that, effectively, there is no bound from low-energy FC processes on this MI for such heavy gluinos and squarks.} and $\tan\beta\sim60$, 
we would obtain:
  \begin{eqnarray}
  \left({\bf R^{-1}_{d}}\right)^{\mathrm{max}}_{32(L)} & \simeq &  -\frac{\delta\epsilon_{32(L)}\left(1+\epsilon\tan\beta\right)\tan\beta}
{\mathrm{Det\left({\bf R}_d\right)}} \sim 
\frac{3 \cdot10^{-3}\,\times\,1.5\,\times\,60}{4} \sim 0.03  
  \label{eq:eqn525}
  \end{eqnarray}
  \vspace{0.1cm}
where $\epsilon$, Eq.~(\ref{eq:eqn505}), and $\mathrm{Det({\bf R}_d)}$ 
for the masses specified above take the values $\epsilon\simeq 0.01$ 
and ${\rm Det}({\bf R}_d)\simeq 4.7$ . Therefore:
  {\small
  \begin{eqnarray}
  \mathrm{BR}\left(H_1 \rightarrow \bar{b}s+\bar{s}b\right)^{\mathrm{max}}_{(L)}
& \simeq & \left| \left({\bf V^{\dagger} R^{-1}_d V}\right)_{32(L)} \right|^2 \frac{3\,\kappa_{QCD} \times (\delta_1^2+\eta_1 ^2)}{\quad \left(3\,\kappa_{QCD}+\frac{m^2_{\tau}}{m_b^2}\right)+I_{PS}\frac{126^2}{m_b^2}\quad } \sim 5 \cdot 10^{-4}\, (\delta_1^2 + \eta_1^2) \label{eq:eqn526} \\
  & & \nonumber \\
  \mathrm{BR}\left(H_2 \rightarrow \bar{b}s+\bar{s}b\right)^{\mathrm{max}}_{(L)}
& \simeq & \left| \left({\bf V^{\dagger} R^{-1}_d V}\right)_{32(L)} \right|^2 \frac{3\,\kappa_{QCD}}{\quad 3\,\kappa_{QCD}+\frac{m^2_{\tau}}{m_b^2}\quad } \sim 10^{-3}\label{eq:eqn527}
  \end{eqnarray}}
\begin{center}
\begin{figure}
\includegraphics[scale=0.4]{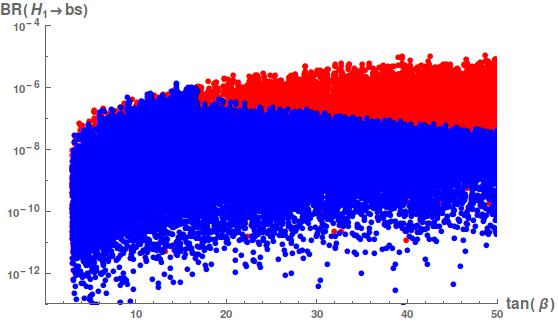}
\includegraphics[scale=0.4]{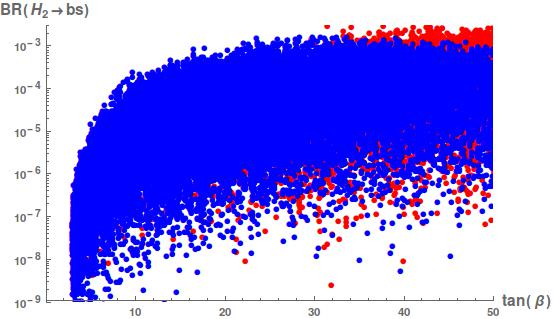}
\includegraphics[scale=0.4]{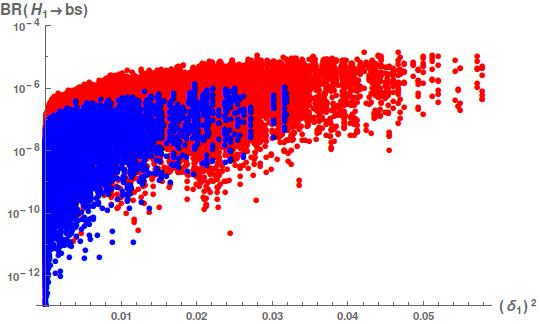}
\includegraphics[scale=0.4]{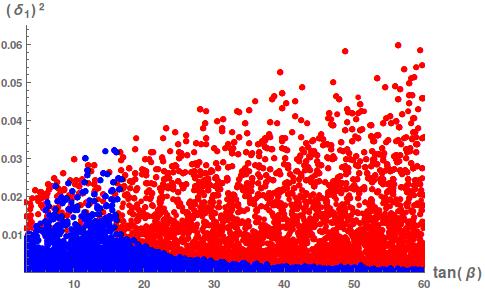}
\caption{
A full MSSM framework with LL insertion  
with $\delta_{L}\neq0$ and $\delta_{A_u}=\delta_{R}=0$:
The upper frames show the dependence of the 
estimated branching ratios for $H_{1,2}\rightarrow \bar{b}s+\bar{s}b$ on 
$\tan\beta$. The lower-left frame is for the dependence of ${\rm B}(H_1\to b s)$ on 
$\delta_1^2$ and the lower-right frame for the $\delta_1^2$ dependence on
$\tan\beta$. Blue (dark) points satisfy all the constraints considered, while red (light) points violate one or several of these constraints. }
\label{fig:501}
\end{figure}
\end{center}
Fig.~\ref{fig:501} shows the results of our scans. Blue points are those which satisfy the whole set of constraints while 
red points are excluded because of the violation of one or more of them. 
In the upper frames of Fig.~\ref{fig:501},
we represent the branching ratios for $H_1$ and $H_2$ versus $\tan\beta$ 
(for all considerations, $H_3$ will be equivalent to $H_2$ given that they are 
nearly degenerated for the range of masses considered here). As can be seen in 
these two plots, before applying the low-energy constraints, the branching ratio 
grows with $\tan\beta$ (red points). However, the final result is very different 
from this when we impose experimental limits from B-mesons (blue points). Whereas 
for heavy Higgses few points become excluded, in the case of  the lightest Higgs 
the effect is more notable. Indeed, looking at the
upper-left frame of Fig.~\ref{fig:501} we can say that the 
tendency is completely opposite and the branching ratio decreases 
for {\small $\tan\beta>20$}.

This behaviour is mainly due to the $B^0_s\rightarrow \mu^+\mu^-$ constraint. This branching ratio is given in Eq.~(\ref{eq:eqn3b9}) with $C_S$, Eq.~(\ref{eq:eqn3b5}), and $C_P$, Eq.~(\ref{eq:eqn3b6}), containing the SUSY contributions.
The dominant contributions come from the heavy Higgses and we have:
  \begin{eqnarray}
  C_{S,P} & \propto & 2\,\frac{\tan^2\beta}{m_{H_j}^2}\,
\left( {\bf V}^{\dagger} {\bf R}^{-1}_d {\bf V}\right)^*_{32}\,
\left[\left(\mathcal{O}_{1j}-\frac{\mathcal{O}_{2j}}
{\tan^2\beta}\right)^2+\mathcal{O}_{3j}^2\right]\:
\sim \: 2\,\frac{\tan^2\beta}{m_{H_j}^2}\,
\left( {\bf R}^{-1}_d\right)^*_{32} \label{eq:eqn528}
  \end{eqnarray}
 with $j=2,3$. This dependence on $\tan^3\beta$ (the matrix element {\small $\left( {\bf R}^{-1}_d\right)^*_{32}$} carries an additional $\tan\beta$) and the heavy Higgs masses explains why this decay 
provides 
such a restrictive constraint for relatively small $m_{H_2,H_3}^2$
and medium-to-large values of $\tan\beta$.

   In fact, BR$(H_1\to bs)$ is suppressed at medium and large $\tan\beta$ values while BR$(H_2\to bs)$ is not. This is due to fact that BR$(H_1\to bs)$ is proportional to $\delta_1$ and $\eta_1$, which in the MSSM are of order $v^2/M_{H_2}^2$. Then, B-meson constraints are more restrictive for light 
$m_{H_2}$
which also correspond to the largest BR$(H_1\to bs)$. On the other hand BR$(H_2\to bs)$ is independent of $\delta_1$ or $\eta_1$ and the $H_2$ mass. The B-meson constraints are not effective here if 
$m_{H_2}$
is large enough and therefore we can reach large branching ratios for large $\tan \beta$ values. Furthermore, this branching ratio saturates for medium-to-large values of $\tan \beta$ when both the FC decay width and the total decay width have the same $\tan \beta$ dependence.  

Moreover, the lower-left frame of Fig.~\ref{fig:501} 
shows the branching ratio for $H_1$ versus $\delta_1^2$. 
As seen in Eq.~(\ref{eq:eqn526}), for 
${\delta_1^2} \gtrsim 0.02$ the branching ratio is of the order $10^{-5}$. However, the implementation of B-meson constraints here reduces this value by more than one order of magnitude.

Finally,  in the lower-right frame of Fig.~\ref{fig:501},
we present the $\delta^2_1$ dependence on $\tan\beta$. 
We observe that, before imposing B-meson constraints, larger values of $(\delta_1^2)$ are possible when $\tan \beta$ grows. 
However, B-meson constraints become very effective for 
$\tan \beta \gsim 17$ as shown by the blue points in this plot.
This could be understood by noting that the 
Higgs contributions to the $\Delta B=1$ and $\Delta B=2$ processes 
are inversely proportional to the 
heavy Higgs mass squared. Therefore, to suppress these processes
for large $\tan\beta$, large Higgs masses are required. 
Accordingly, we expect smaller $|\delta_1|$ as $\tan \beta$ grows
since, as we have seen, $|\delta_1| \propto v^2/m_{H_2}^2$ in the MSSM.

\begin{center}
\begin{figure}[t!]
\includegraphics[scale=0.37]{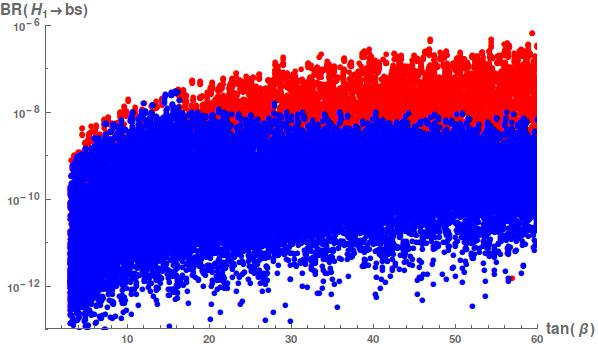}
\includegraphics[scale=0.37]{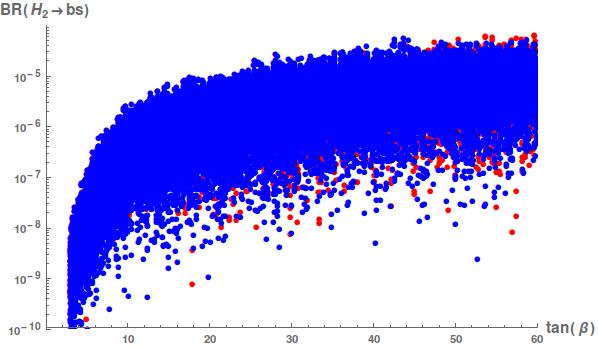}
\caption{
A full MSSM framework with RR insertion  
with $\delta_{R}\neq0$ and $\delta_{A_u}=\delta_{L}=0$:
The left frame shows the dependence of ${\rm BR}(H_1\to b s)$ on 
 $\tan\beta$ and the right frame shows
the dependence of ${\rm BR}(H_2\to b s)$ on $\tan\beta$. Again, blue (dark) points satisfy all the constraints considered, while red (light) points violate one or several of these constraints.
}
\label{fig:504}
\end{figure}
\end{center}
\subsubsection {Right-handed (R) insertion}
Now we consider the case, 
$\delta_{R}\neq 0$ and $\delta_L=\delta_{A_u}=0$, 
we have from Eqs.~(\ref{eq:eqn508}) and (\ref{eq:eqn524}): 
  \begin{eqnarray}
  \delta\epsilon_{32(R)} & \simeq & 
\delta_R
\left(\frac{{\bf V}^{*}_{22}\,y_s}{{\bf V}^{*}_{33}\,y_b}\,-\,\frac{{{\bf V}^{*}_{23}}^2\,y_b}{{\bf V}^{*}_{22}\,{\bf V}^*_{33}\,y_s}\right)\, 
\frac{\delta\epsilon_{32(L)}}{\delta_L} \sim
 0.013 \times 3 \times 10^{-3}\, \delta_{R}  \sim 5\cdot 10^{-5}\, \delta_{R}\,, \\
  &  & \nonumber \\
  \left({\bf R}^{-1}_{d}\right)_{32(R)} & \simeq & \left(\frac{{\bf V}^{^*}_{22}\,y_s}{{\bf V}^{^*}_{33}\,y_b}\,-\,\frac{{{\bf V}^{*}_{23}}^2\,y_b}{{\bf V}^{^*}_{22}\,{\bf V}^{^*}_{33}\,y_s}\right)\left({\bf
R}^{-1}_{d}\right)_{32(L)} \sim 4 \cdot 10^{-4}\, \delta_{R} \,.
  \end{eqnarray}
  \vspace{0.01cm}
In this case, the value of the off-diagonal element $({\bf R}^{-1}_{d})_{32}$ is of the order of $10^{-4}$ and this implies that the contributions ${\bf V}^*_{23}{\bf V}_{22}$ in Eq.~(\ref{eq:eqn522}) and {\small $\left({\bf
R}^{-1}_d\right)_{23}$} in Eq.~(\ref{eq:eqn521}) or $y_{R_i}$ in Eq.~(\ref{eq:eqn512}), can be important. Thus:
  \begin{eqnarray}
  \left( {\bf V}^\dagger {\bf R}^{-1}_d 
{\bf V} \right)^{\mathrm{max}}_{32(R)} 
& \simeq & \left({\bf R}^{-1}_{d}\right)_{32(R)}\, +\, 
2\cdot 10^{-3} \left({\bf R}^{-1}_{d}\right)_{23(R)}+\, 
0.04\, \frac{\left(1+\epsilon\tan\beta\right)\eta\,\tan\beta}
{\mathrm{Det\left({\bf R}_d\right)}}\nonumber \\
  & \sim & -4\cdot 10^{-4} \delta_{R} -2\cdot 10^{-3}\, \frac{\delta\epsilon_{23(R)}\times 1.5\times 60}{4}  + 0.04\, \frac{1.5\times 4\cdot10^{-3} \times 60}{4}
  \label{eq:eqn528}
  \end{eqnarray}
where we used the same values for the parameters as in Eq.~(\ref{eq:eqn525}) and $\delta\epsilon_{23(R)}$ is now enhanced by a factor $y_b/y_s$:
{\small
  \begin{eqnarray}
  \delta\epsilon_{23(R)} & \sim & \delta_R\frac{2\alpha_s}{3\pi}\, \frac{y_b}{y_s}\, \mu^*\, M_3^*\, \tilde{M}^2_{D_3}\, K\left(\tilde{M}^2_Q, \tilde{M}^2_D, \left| M_3 \right|^2 \right) \sim -0.1\,\delta_{R}\,.
  \end{eqnarray}}
Therefore, we obtain,
  \begin{eqnarray}
 \left( {\bf V}^{\dagger} {\bf R}^{-1}_d {\bf V}\right)^{\mathrm{max}}_{32(R)}  & \sim & 4(\delta_{R} + 1)\cdot 10^{-3}
\end{eqnarray}    
Thus, for $\delta_{R}=0.5$, the branching ratios are:
  {\small
  \begin{eqnarray}
  \mathrm{BR}\left(H_1 \rightarrow \bar{b}s+\bar{s}b\right)^{\mathrm{max}}_{(R)} & \simeq & \left| \left( {\bf V}^{\dagger} {\bf R}^{-1}_d {\bf V}\right)_{32(R)} \right|^2 \frac{3\,\kappa_{QCD} (\delta_1^2 + \eta_1^2)}{\quad\left(3\,\kappa_{QCD}+\frac{m^2_{\tau}}{m_b^2}\right)+I_{PS}\frac{126^2}{m_b^2}\quad } \sim 1.5  \cdot 10^{-5}\,	 (\delta_{1}^2 + \eta_1^2) \label{eq:eqn529}\\
  & & \nonumber \\
  \mathrm{BR}\left(H_2 \rightarrow \bar{b}s+\bar{s}b\right)^{\mathrm{max}}_{(R)} & \simeq & \left| \left( {\bf V}^{\dagger} {\bf R}^{-1}_d {\bf V}\right)_{32(R)} \right|^2 \frac{3\,\kappa_{QCD}}{\quad 3\,\kappa_{QCD}+\frac{m^2_{\tau}}{m_b^2}\quad } \sim 2\cdot 10^{-5} \label{eq:eqn530}
  \end{eqnarray}}

The results of our scans for this case are shown in  
Fig.~\ref{fig:504}. As before, these results are in agreement with the numerical 
values if B-meson constraints are not taken into account. 
Once they are incorporated into the analysis, the lightest Higgs branching 
ratio is reduced by more than one order of magnitude. Also,  taking into account, 
from Eq.~(\ref{eq:eqn528}), that 
both the MI and MI-independent contributions
are of the same order only for large $\delta_{R}$, 
the BR is completely independent of $\delta_{R}$.

\begin{center}
\begin{figure}[t!]
\includegraphics[scale=0.4]{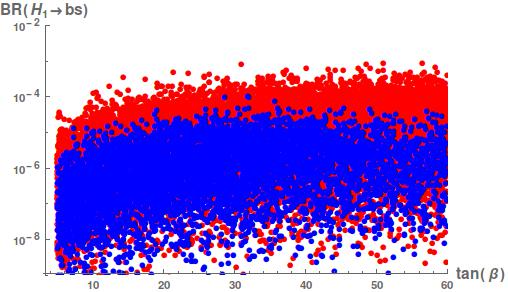}
\includegraphics[scale=0.4]{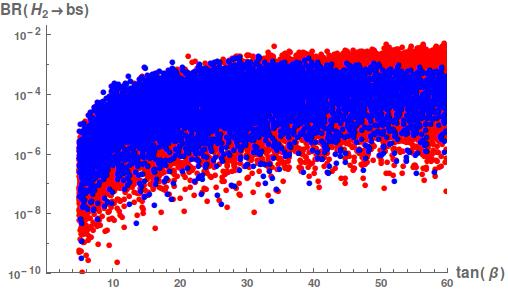}
\includegraphics[scale=0.4]{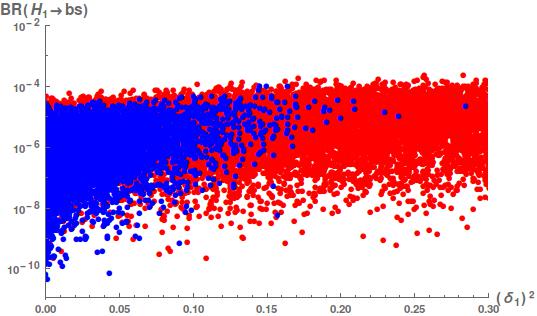}
\includegraphics[scale=0.4]{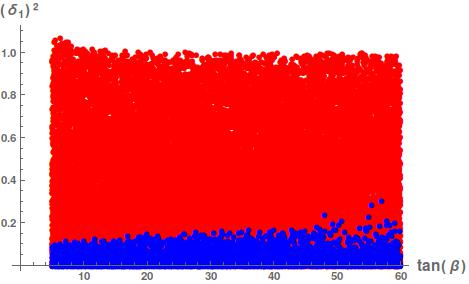}
\caption{
A generic supersymmetric SM with L insertion
with $\delta_{L}\neq0$ and $\delta_{A_u}=\delta_{R}=0$:
The upper frames show the dependence of the 
estimated branching ratios for $H_{1,2}\rightarrow \bar{b}s+\bar{s}b$ on 
$\tan\beta$. The lower-left frame is for the dependence of ${\rm B}(H_1\to b s)$ on 
$\delta_1^2$  and the lower-right frame for the $\delta_1^2$ dependence on
$\tan\beta$. Blue (dark) points satisfy all the constraints considered, while red (light) points violate one or several of these constraints.}
\label{fig:506}
\end{figure}
\end{center}
\subsection{Generic supersymmetric SM}
After computing these branching ratios in the MSSM framework, we now perform our analysis in a generic supersymmetric model. Therefore, we present here a model-independent analysis, meaning that, in fact, we consider a generic Higgs mixing matrix with possible additional Higgs states. In this case, we have
$\mathcal{O}_{11}^2+\mathcal{O}_{21}^2+\mathcal{O}_{31}^2\leq1$ and the parameters
$\delta_1$ and/or $\eta_1$ entering $\Gamma (H_1 \to \bar b s + \bar s b)$ can be sizeable.

The expressions for the decay widths and branching ratios, 
Eqs.~(\ref{eq:eqn513}--\ref{eq:eqn520}),
are still valid in the generic supersymmetric scenario. 
The main difference now is that the parameters $\delta_1$ and 
$\eta_1$ are only constrained by experimental results on Higgs decays and low-energy FCNC processes. 
Notice, however, that 
the flavour-changing entries in ${\bf R^{-1}_{d}}$ do not change in the 
two models.  

In Fig.~\ref{fig:506},
we show the FC branching ratios of $H_1$ and $H_2$ 
for $\delta_{L}\neq0$ and $\delta_{A_u}=\delta_{R}=0$ in the 
generic supersymmetric SM. 
These figures can be compared with Fig.~\ref{fig:501} which 
shows the corresponding branching ratios in the MSSM framework. 

In the MSSM framework, the mixing angles are obtained through a minimization of the scalar potential and both $\delta_1$ and $\eta_1$ are of the order of 
$v^2/m_{H_2}^2$
from the diagonalization of the neutral Higgs mass matrix. In the generic supersymmetric scenario we treat $\delta_1$ and $\eta_1$ as free parameters that do not depend a priori on the ratio $v^2/m_{H_2}^2$,
but are only constrained by the different experimentally measured Higgs branching ratios and B-meson constraints. This is why BR$(H_1 \to bs)$ in Fig~\ref{fig:501} is two orders of magnitude smaller that the largest possible value in Figure~\ref{fig:506}. The different distribution of the points allowed by B-meson constraints is due to the same reason. In the MSSM scenario, Higgs flavour changing processes are mediated by the heavy Higgses and therefore are only important for light $m_{H_2}$
which, as we have seen, correspond also to the largest $\delta_1$ and 
$\eta_1$ and therefore to the largest branching ratios.
In the generic supersymmetric SM, it is, in principle, possible to have a large $\delta_1$ with a heavy $H_2$ and therefore the B-meson constraints are not so efficient. 

On the other hand, the FC decays of the heavy Higgses $H_{2,3}$ are independent of the values of $\delta_1$ and $\eta_1$ as can be seen in 
Eq.~(\ref{eq:eqn520}). 
Therefore,  the upper-right frames of
Figs.~\ref{fig:501} and \ref{fig:506}
are very similar and we obtain very similar results for BR$(H_2 \to bs)$ in the MSSM framework and in the generic supersymmetric SM.

Also, as shown in the lower-right plot in Fig.~\ref{fig:506}, the allowed values of $(\delta_1)^2$ are completely independent of $\tan \beta$ as the B-meson constraints, which depending on $(\tan \beta)^n/m_{H_2}^4$ can always be satisfied by adjusting conveniently the value of $m_{H_2}$. In this case, the upper limit for $(\delta_1)^2$ is fixed by the $H_1 \to \gamma \gamma$ decay which as shown in \cite{Barenboim:2013qya}, requires $\left({\cal O}^2_{11}+{\cal O}^2_{31}\right)\sim 1/\tan^2\beta$ and ${\cal O}^2_{21} \sim 1-1/\tan^2 \beta$. Using the definition of $\delta_1$ and in the limit ${\cal O}_{31}\ll 1$, with the above constraints, $(\delta_1)^2 \lsim 0.17$ for $\tan \beta\gtrsim 10$, as we see numerically in this plot.   

In summary, the main difference in the generic supersymmetric SM is that   BR$(H_1 \to bs)$ could reach a value of $\sim 10^{-4}$ consistently with present experimental constraints. This value is still too small to be observed in the large background of a hadron collider, but it could be tested in a leptonic linear collider in the near future.        

\section{Conclusions} \label{sec:conclusions}

In this paper, we have analyzed the FC Higgs decay $H_i \to b s$ for the different Higgs states in both an MSSM scenario and a more general supersymmetric SM framework. The importance of this observable is that effects of heavy particles  do not decouple and may provide a first sign of new physics for heavy supersymmetric masses beyond collider reach. 
Before we carried out our  numerical analysis,
we derived approximated analytic expressions for the
off-diagonal entries of
$\left({\bf V}^\dagger {\bf R}_d^{-1} {\bf V}\right)$
which dictate the size of flavour violation.
In an MSSM framework we showed that, even in the presence of large off-diagonal flavour entries in the sfermion mass matrices, for the light Higgs,  BR$(H_1 \to b s) \lesssim 10^{-6}$ consistently with present experimental constraints, while for heavy Higgs states  BR$(H_{2,3} \to b s)$ can still be $\sim 10^{-3}$. In a more general supersymmetric scenario, where we allowed for non-minimal Higgs mixings, the branching ratio BR($H_1 \to b s$) can reach values $\sim {\cal
O}(10^{-4})$, while BR$(H_{2,3} \to b s)$ remain of the order of $\sim 10^{-3}$. 
We find that the results of the numerical analysis are well in accord with
the estimations made using the approximated analytic expression for
$\left({\bf V}^\dagger {\bf R}_d^{-1} {\bf V}\right)$.
Although these small branching ratios
are clearly out of reach for the LHC due to the very large $b$-quark background, a full study in a linear collider environment could still be worth pursuing.    

\section*{Acknowledgments}
G.B., C.B,  M.L.L.I and O.V. acknowledge support from the MEC and FEDER (EC)
Grants FPA-2011-23596 and FPA2014-54459-P and the Generalitat Valenciana under grant  PROMETEOII/2013/017. G.B. acknowledges partial support from the European Union FP7 ITN INVISIBLES (Marie Curie Actions, PITN-GA-2011-289442). C.B. thanks {\it Ministerio de Educaci\'on, Cultura y Deporte} for financial support through an FPU-grant AP2010-3316. The work of M.L.L.I is funded through an FPI-grant   BES-2012-053798  from {\it Ministerio de Economia y Competitividad}. 
J.S.L. was supported by
the National Research Foundation of Korea (NRF) grant
(No. 2013R1A2A2A01015406).

\appendix
\section{FC Higgs couplings} \label{sec:appendixa}
In this appendix, we present the explicit expression for {\small ${\bf G}_d^0$} associated with the FCNC Higgs couplings in Eq.~(\ref{eq:eqn213}),
\begin{equation}
  \hspace{-11.5cm}
  {\bf G}^0_d\,=\,\langle{\bf \Delta}^{\Phi_2}_d\,+\,{\bf \delta \Delta}^{\Phi_2}_d\rangle_{_0}
  \label{eq:eqn219}
  \end{equation}
  \begin{eqnarray}
  \langle{\bf \Delta}^{\Phi_2}_d\rangle_{_0} & = & {\bf 1}\,\frac{2\alpha_3}{3\pi}\,\mu^*\,M_3^*\,I\left(\widetilde{M}^2_Q,\widetilde{M}^2_D,\left|M_3\right|^2\right)\,-\,{\bf 1}\,\frac{\alpha_1}{36\pi}\mu^*\,M_1^*\,I\left(\widetilde{M}^2_Q,\widetilde{M}^2_D,\left|M_1\right|^2\right) \nonumber \\
  & & \hspace{-0.15cm} +\,\frac{{\bf h}^{\dagger}_u{\bf h}_u}{16\pi^2}\,\mu^*\,A^*_u\,I\left(\widetilde{M}^2_Q,\widetilde{M}^2_U,\left|\mu\right|^2\right)\,-\,\frac{3\alpha_2}{8\pi}\mu^*\,M_2^*\,I\left(\widetilde{M}^2_Q,\widetilde{M}^2_D,\left|\mu\right|^2\right) \nonumber \\
  & & \hspace{-0.15cm} -\,{\bf 1}\,\frac{\alpha_1}{24\pi}\,\mu^*\,M_1^*\,I\left(\widetilde{M}^2_Q,\left|M_1\right|^2,\left|\mu\right|^2\right)-\,{\bf 1}\,\frac{\alpha_1}{12\pi}\,\mu^*\,M_1^*\,I\left(\widetilde{M}^2_D,\left|M_1\right|^2,\left|\mu\right|^2\right) \label{eq:eqn220} \\
  & & \nonumber \\
  \langle{\bf \delta \Delta}^{\Phi_2}_d\rangle_{_0} & = & \frac{2\alpha_3}{3\pi}\,\mu^*\,M_3^*\left[{\bf \delta}\widetilde{{\bf M}}_Q^2\,K\left(\widetilde{M}^2_Q,\widetilde{M}^2_D,\left|M_3\right|^2\right)\,+\,{\bf h}_d^{-1}{\bf \delta}\widetilde{{\bf M}}_D^2{\bf h}_d\,K\left(\widetilde{M}^2_D,\widetilde{M}^2_Q,\left|M_3\right|^2\right)\right] \nonumber \\
  & & \hspace{-0.35cm} -\,\frac{\alpha_1}{36\pi}\mu^*\,M_1^*\,\left[{\bf \delta}\widetilde{{\bf M}}_Q^2\,K\left(\widetilde{M}^2_Q,\widetilde{M}^2_D,\left|M_1\right|^2\right)\,+\,{\bf h}^{-1}_d{\bf \delta}\widetilde{{\bf M}}_D^2{\bf h}_d\,K\left(\widetilde{M}^2_D,\widetilde{M}^2_Q,\left|M_1\right|^2\right)\right] \nonumber \\
  & & \hspace{-0.45cm} +\,\frac{1}{16\pi^2}\,\mu^*\,A^*_u\,\left[{\bf h}_u^{\dagger}{\bf \delta}\widetilde{{\bf M}}_U^2{\bf h}_u\,K\left(\widetilde{M}^2_U,\widetilde{M}^2_Q,\left|\mu\right|^2\right)\,+\,
{\bf \delta}\widetilde{\bf M}_Q^2{\bf h}_u^{\dagger}{\bf h}_u\,
K\left(\widetilde{M}^2_Q,\widetilde{M}^2_U,\left|\mu\right|^2\right)\right] \nonumber \\ 
  & & \hspace{-0.45cm}+\,\frac{{\bf \delta a}_u^{\dagger}{\bf h}_u}{16\pi^2}\,\mu^*\,I\left(\widetilde{M}^2_Q,\widetilde{M}^2_U,\left|\mu\right|^2\right)\,-\,\frac{3\alpha_2}{8\pi}\,\mu^*M_2^*{\bf \delta}\widetilde{\bf M}_Q^2\,K\left(\widetilde{M}_Q^2,\left|\widetilde{M}_2\right|^2,\left|\mu\right|^2\right) \label{eq:eqn221} \\
  & & \hspace{-0.45cm} -\,\frac{\alpha_1}{24\pi}\,\mu^*\,M_1^*{\bf \delta}\widetilde{\bf M}_Q^2\,K\left(\widetilde{M}^2_Q,\left|M_1\right|^2,\left|\mu\right|^2\right)\,-\,\frac{\alpha_1}{12\pi}\,\mu^*\,M_1^*{\bf h}_d^{-1}
{\bf \delta}\widetilde{\bf M}_D^2{\bf h}_d\,
K\left(\widetilde{M}^2_D,\left|M_1\right|^2,\left|\mu\right|^2\right) \nonumber
  \end{eqnarray}
\vspace{0.25cm}  
where the loops functions are given by:
  \vspace{0.1cm}
  \begin{eqnarray}
  I\left(a,b,c\right) & = & \frac{ab\,\ln{\left(a/b\right)}\,+\,bc\,\ln{\left(b/c\right)}\,+\,ac\,\ln{\left(c/a\right)}}{(a-b)(b-c)(a-c)} \label{eq:eqn222}\\
  & & \nonumber \\
  K\left(a,b,c\right) & = & \frac{\mathrm{d}}{\mathrm{da}}\Biggl[I\left(a,b,c\right)\Biggr]\,=\, \frac{b\ln{\left(a/b\right)}\,+\,c\ln{\left(c/a\right)}}{(a-b)(b-c)(a-c)}\,+\,\frac{(b+c-2a)\,I\left(a,b,c\right)+1}{(a-b)(a-c)} \label{eq:eqn223}
  \end{eqnarray}
\vspace{0.1cm}  
From here, the elements $\epsilon$, $\delta$, used in the {\small ${\bf G}_d^0$} matrix in Eq.~(\ref{eq:eqn504}) are, 
{\small
  \begin{eqnarray}
  \epsilon & = & \frac{2\alpha_s}{3\pi}\, \mu^*\, M_3^*\,I\left( \tilde{M}^2_Q,\tilde{M}^2_D,\left| M_3 \right|^2 \right)\,+\,\frac{\rho-1}{3}\left[\frac{2\alpha_s}{3\pi}\, \mu^*\, M_3^* \left(\tilde{M}^2_{D_3}+\tilde{M}^2_{Q_3}\right)\, K\left(\tilde{M}^2_Q, \tilde{M}^2_D, \left| M_3 \right|^2 \right) \right]   \label{eq:eqn505} \\
       &   &  \nonumber \\     
  \eta & = & \frac{\left|y_t\right|^2}{16\pi^2}\,\mu^* A_u^*\,I\left(\tilde{M}^2_Q, \tilde{M}^2_U, \left| \mu \right|^2\right)\,-\,\delta_{R}\left[\frac{2\alpha_s}{3\pi}\, \mu^*\, M_3^*\,\frac{V^{^*}_{23}}{V^{^*}_{33}}\tilde{M}^2_{D_3}\, K\left(\tilde{M}^2_Q, \tilde{M}^2_D,\left| M_3 \right|^2 \right)\right] \label{eq:eqn506} \\
       &   & \hspace{-0.4cm} +\,\left(1-\rho\right)\left[\frac{2}{3}\frac{\left|y_t\right|^2}{16\pi^2}\,\mu^*\,\Biggl(A^*_t\,I\left(\tilde{M}^2_Q, \tilde{M}^2_U, \left| \mu \right|^2\right)\,+\, A_u^*\left(\tilde{M}^2_{U_3}+\tilde{M}^2_{Q_3}\right)\, K\left(\tilde{M}^2_Q, \tilde{M}^2_U, \left| \mu \right|^2\right)\Biggr) \right. \nonumber \\
       &   & \hspace{-0.4cm} \left.+\,\frac{2\alpha_s}{3\pi}\, \mu^*\, M_3^*\,\left(\tilde{M}^2_{D_3}+\tilde{M}^2_{Q_3}\right)\,K\left(\tilde{M}^2_Q, \tilde{M}^2_D, \left| M_3 \right|^2 \right) \right] \nonumber
  \end{eqnarray}}
\section{B-physics constraints} \label{sec:appendixb}
The main FC processes associated with B-Mesons that we consider in our analysis are $\Delta M_{B_s}$ and $\bar{B}^0_s\rightarrow\mu^+\mu^-$, although other constraints like BR$(B \to X_s \gamma)$ are also included.

In the case of $\Delta M_{B_s}$, we use the expression in \cite{Ellis:2007kb}, given by:
\begin{equation}
\Delta M_{B_s}=2\left|\left\langle\bar{B}^0_s|H^{\Delta
B=2}_{eff}|B^0_s\right\rangle_{SM}\, +\, \left\langle\bar{B}^0_s|H^{\Delta
B=2}_{eff}|B^0_s\right\rangle_{SUSY}\right|\
\label{eq:eqn3a1}
\end{equation}
where the SUSY contribution is \cite{Ellis:2007kb}:
{\small
\begin{eqnarray}
\left\langle\bar{B}^0_s|H^{\Delta B=2}_{eff}|B^0_s\right\rangle_{SUSY} & = & 2310\mathrm{ps}^{-1}\left(\frac{\widehat{B}^{1/2}_{B_s}F_{B_s}}{265\,\mathrm{MeV}}\right)^2\left(\frac{\nu_B}{0.55}\right)\times \nonumber \\ 
&  & \left[0.88\left(C_2^{LR(DP)}+C_2^{LR(2HDM)}\right)
-0.52\left(C_1^{SLL(DP)}+C_1^{SRR(DP)}\right)\right] \label{eq:eqn3a2}\,.
\end{eqnarray}}
where the Wilson coefficients above $C_2^{LR(DP)}$, $C_2^{LR(2HDM)}$, $C_1^{SLL(DP)}$ and $C_1^{SRR(DP)}$ are associated with double-penguin and box diagrams.

The Wilson coefficients $C_1^{SLL(DP)}$, $C_1^{SRR(DP)}$, $C_2^{LR(DP)}$ and $C_2^{LR(2HDM)}$ related to the SUSY contribution of the $B_s$-meson mass difference in Eq.~(\ref{eq:eqn3a2}) are, 
\begin{eqnarray}
C_1^{SLL(DP)} & = & -\frac{16\pi^2m_b^2}{\sqrt{2}G_FM^2_W}\sum^{3}_{i=1}\frac{{\bf g}^L_{H_i\bar{b}s}\,{\bf g}^L_{H_i\bar{b}s}}{m_{H_i}} \label{eq:eqn3a3} \\
C_1^{SRR(DP)} & = & -\frac{16\pi^2m_s^2}{\sqrt{2}G_FM^2_W}\sum^{3}_{i=1}\frac{{\bf g}^R_{H_i\bar{b}s}\,{\bf g}^R_{H_i\bar{b}s}}{m_{H_i}} \label{eq:eqn3a4} \\
C_2^{LR(DP)} & = & -\frac{32\pi^2m_bm_s}{\sqrt{2}G_FM^2_W}\sum^{3}_{i=1}\frac{{\bf g}^L_{H_i\bar{b}s}\,{\bf g}^R_{H_i\bar{b}s}}{m_{H_i}} \label{eq:eqn3a5} \\ 
C_2^{LR(2HDM)} & = & \left.C_{2}^{LR(2HDM)}\right|_{H^{\pm}H^{\mp}}\:+\:\left.C_{2}^{LR(2HDM)}\right|_{W^{\pm}H^{\mp}}\,.
\label{eq:eqn3a6}
\end{eqnarray}
\vspace{0.1cm}
The couplings of the charged Higgses and Goldstone bosons to fermions, Eq.~(3.38) of Ref.\cite{Ellis:2009di}, are given by:
\begin{eqnarray}
\mathcal{L} & \supset & -\frac{g}{2M_W}\left[
H^{-}\bar{d}\left(\widehat{{\bf M}}_d\,{\bf g}_{H^-}^L\,P_L+
{\bf g}_{H^-}^R\,\widehat{\bf M}_u\,P_R\right)u
\,+\,G^- \bar{d} \left(\widehat{\bf M}_d\, {\bf g}^L_{G^-}\,P_L +
{\bf g}^R_{G^-}\,\widehat{\bf M}_u\,P_R\right) u
\right] +{\rm h.c.}\,
\label{eq:eqn3a7}
\end{eqnarray}
and in the large $\tan\beta$ limit, we have:
\begin{equation}
\begin{array}{lll}
\mathrm{\textbf{g}}^L_{H^-}=-\tan\beta\mathrm{\textbf{V}}^{\dagger}\mathrm{\textbf{R}}^{-1}_{d}+
\mathrm{\textbf{V}}^{\dagger}\mathrm{\textbf{R}}^{-1}_{d}\mathrm{\textbf{G}}^0_d & \hspace{1.cm}  & \mathrm{\textbf{g}}^{R}_{H^-}=-\frac{1}{\tan\beta}\mathrm{\textbf{V}}^{\dagger}
\end{array}
\label{eq:eqn3a8}
\end{equation}
\begin{equation}
\begin{array}{lll}
\hspace{-0.7cm} \mathrm{\textbf{g}}^{L}_{G^-}=\mathrm{\textbf{V}}^{\dagger} & \hspace{1.cm}  &  \hspace{4.3cm} \mathrm{\textbf{g}}^{R}_{G^-}=-\mathrm{\textbf{V}}^{\dagger}
\end{array}
\label{eq:eqn3a9}
\end{equation}
\vspace{0.1cm}
Then $C_2^{LR(2HDM)}$ includes two main contributions: one associated with box diagrams for two $H^{\pm}_l$ and another for $W^{\mp}H^{\pm}$ box diagrams. From Eqs.~(4.4) and (4.5) in Ref.\cite{Buras:2001mb}, we have:
\begin{eqnarray}
\left.C_{2}^{LR(2HDM)}\right|_{H^{\pm}H^{\mp}} & = & \frac{8m_bm_sm_t^4}{M^2_W} \sum_{k,l=H,G} {\mathrm{\textbf{g}}^L_{H^-_l}}_{33}{\mathrm{\textbf{g}}^{L\dagger}_{H^-_l}}_{32}{\mathrm{\textbf{g}}^R_{H^-_k}}_{33}{\mathrm{\textbf{g}}^{R\dagger}_{H^-_k}}_{32}D_0\left(M^2_{H^-_l},M^2_{H^-_k},m_t^2,m_t^2\right)
\label{eq:eqn3a10}
\end{eqnarray}
\begin{eqnarray}
\left.C_{2}^{LR(2HDM)}\right|_{W^{\pm}H^{\mp}} & = & -8m_bm_s \sum^{3}_{i,j=1}\sum_{k=H,G} {\mathrm{\textbf{g}}^L_{H^-_k}}_{3i}{\mathrm{\textbf{g}}^{L\dagger}_{H^-_k}}_{j2}{\mathrm{\textbf{V}}^{\dagger}}_{3j}{\mathrm{\textbf{V}}}_{i2}~
D_2\left(M^2_W,M^2_{H^-_k},m_{q_i}^2,m_{q_j}^2\right)
\label{eq:eqn3a12}
\end{eqnarray}
where $D_0(a,b,c,d)$ and  $D_2(a,b,c,d)$ are the corresponding loop functions which can be found in  Ref.\cite{Buras:2001mb}

The decay $\bar{B}^0_s\rightarrow\mu^+\mu^-$ is described by the effective Hamiltonian,
\begin{equation}
H^{\Delta B=1}_{\mathrm{eff}}\,=\,-2\sqrt{2}G_FV_{tb}V^{^*}_{ts}\left(C_S\mathcal{O}_S + C_P\mathcal{O}_P + C_{10}\mathcal{O}_{10}\right)
\label{eq:eqn3b1}
\end{equation}
where the relevant operators are 
$\mathcal{O}_S =  \frac{e^2}{16\pi^2}m_b\left(\bar{q}P_Rb\right)\left(\bar{\mu}\mu\right)$, $\mathcal{O}_P  =  \frac{e^2}{16\pi^2}m_b\left(\bar{q}P_Rb\right)\left(\bar{\mu}\gamma_5\mu\right)$ and $\mathcal{O}_{10}  =  \frac{e^2}{16\pi^2}\left(\bar{q}\gamma^{\mu}P_Lb\right)\left(\bar{\mu}\gamma_{\mu}\gamma_5\mu\right)$.

Neglecting the non-holomorphic vertices on the leptonic sector 
as well as the contributions proportional to the lighter quark masses $m_{d,s}$,
the branching ratio is given by 
\begin{equation}
\mathrm{BR}\left(\bar{B}^0_s\,\rightarrow\,\mu^+\mu^-\right)\,=\,\frac{G_F^2\alpha^2_{\mathrm{em}}}{16\pi^3}\,M_{B_s}\tau_{B_s}\,\left|V_{tb}V^{^*}_{ts}\right|^2\sqrt{1-\frac{4\,m^2_{\mu}}{M^2_{B_s}}}\left[\left(1-\frac{4m^2_{\mu}}{M^2_{B_s}}\right)\left|F_S^s\right|^2+\left|F^s_P\,+\,2m_{\mu}F^{s}_A\right|^2\right]
\label{eq:eqn3b9}
\end{equation}
where $\tau_{B_s}$ is the total lifetime of the $B_s$ meson and the form factors are:
\begin{equation}
F_{S,P}^s\,=\,-\frac{i}{2}\,M^2_{B_s}F_{Bs}\,\frac{m_b}{m_b\,+\,m_q}\,C_{S,P} \,,
\hspace{1.cm} F_A^s=\,-\frac{i}{2}\,F_{B_s}\,C_{10}\,,
\label{eq:eqn3b10}
\end{equation}
with the Wilson coefficients,
\begin{eqnarray}
C_S & = & \frac{2\pi
m_{\mu}}{\alpha_{\mathrm{em}}}\,\frac{1}{V_{tb}V_{ts}^{^*}}\:\sum_{i=1}^{3}\:
\frac{{\bf g}^R_{H_i\bar{s}b}\,g^S_{H_i\bar{\mu}\mu}}{m_{H_i}^2}
\label{eq:eqn3b5}\,, \\
C_P & = & i\,\frac{2\pi
m_{\mu}}{\alpha_{\mathrm{em}}}\,\frac{1}{V_{tb}V_{ts}^{^*}}\: \sum_{i=1}^{3}\:
\frac{{\bf g}^R_{H_i\bar{s}b}\,g^P_{H_i\bar{\mu}\mu}}{m_{H_i}^2}
\label{eq:eqn3b6}\,, \\
C_{10} & = & -4.221 \label{eq:eqn3b7}
\end{eqnarray}
where $C_{10}$ is the leading SM contribution, and $g_{H_i\bar{\mu}\mu}^S\,=\,\frac{\,\mathcal{O}_{1i}\,}{\cos\beta}$ and $g_{H_i\bar{\mu}\mu}^P\,=\,-\tan\beta\,\mathcal{O}_{3i}$ are the Higgs couplings to the charged leptons.

\end{document}